\documentclass[superscriptaddress,amsmath,amssymb,aps]{revtex4-2}

\usepackage{graphicx}
\usepackage{dcolumn}
\usepackage{bm}
\usepackage[pdftex,bookmarks,colorlinks]{hyperref}
\usepackage{lineno}
\usepackage{booktabs}
\usepackage{array}
\usepackage{rotating}
\usepackage{siunitx}

\begin{document}

\preprint{APS/123-QED}

\title{Quantum Networks Using Color Defects in Diamond: Principles, Progress, and Perspectives}

\author{Ayan Majumder}
\affiliation{Department of Electrical Engineering, Indian Institute of Technology Bombay, Mumbai-400076, India}

\author{Cem G\"uney Torun}%
\affiliation{Institut für Physik, Humboldt-Universität zu Berlin, Berlin, Germany}

\author{Tim Schr\"oder}%
\affiliation{Institut für Physik, Humboldt-Universität zu Berlin, Berlin, Germany}
\affiliation{Ferdinand-Braun-Institut (FBH), Berlin, Germany}

\author{Gregor Pieplow}%
\affiliation{Institut für Physik, Humboldt-Universität zu Berlin, Berlin, Germany}

\author{Prem Kumar}%
\affiliation{Center for Photonic Communication and Computing, Department of Electrical and Computer Engineering, Northwestern University, 2145 Sheridan Road, Evanston, IL-60208, USA}

\author{Kasturi Saha}%
\email{kasturis@iitb.ac.in}
\affiliation{Department of Electrical Engineering, Indian Institute of Technology Bombay, Mumbai-400076, India}
\affiliation{Center of Excellence for Quantum Information, Computing Science and Technology, Indian Institute of Technology Bombay, Mumbai-400076, India.}



\begin{abstract} 

Large-scale quantum networks will enable entirely new applications of quantum information science in fields such as quantum communication, distributed quantum computing, sensing, and metrology. To build nodes of such networks, diamond color defects are one of the promising candidates. Their excellent optical properties, fast spin-qubit control, and long spin coherence times make them well-suited for quantum information processing and quantum memory applications. Additionally, recent advances in the heterogeneous integration of diamond nanophotonic structures with photonic integrated circuits have made these systems more efficient and well-suited for scalable quantum processor architectures. In this comprehensive review, we discuss the optical and spin properties of these systems, recent progress in the building blocks of quantum networks, and demonstrations of metropolitan-scale quantum networks, as well as the challenges associated with these systems at both the fundamental and experimental levels, along with potential solutions.

\end{abstract}

\maketitle
\section{Introduction}

\begin{figure*}[]
\centering
\includegraphics[width=\linewidth]{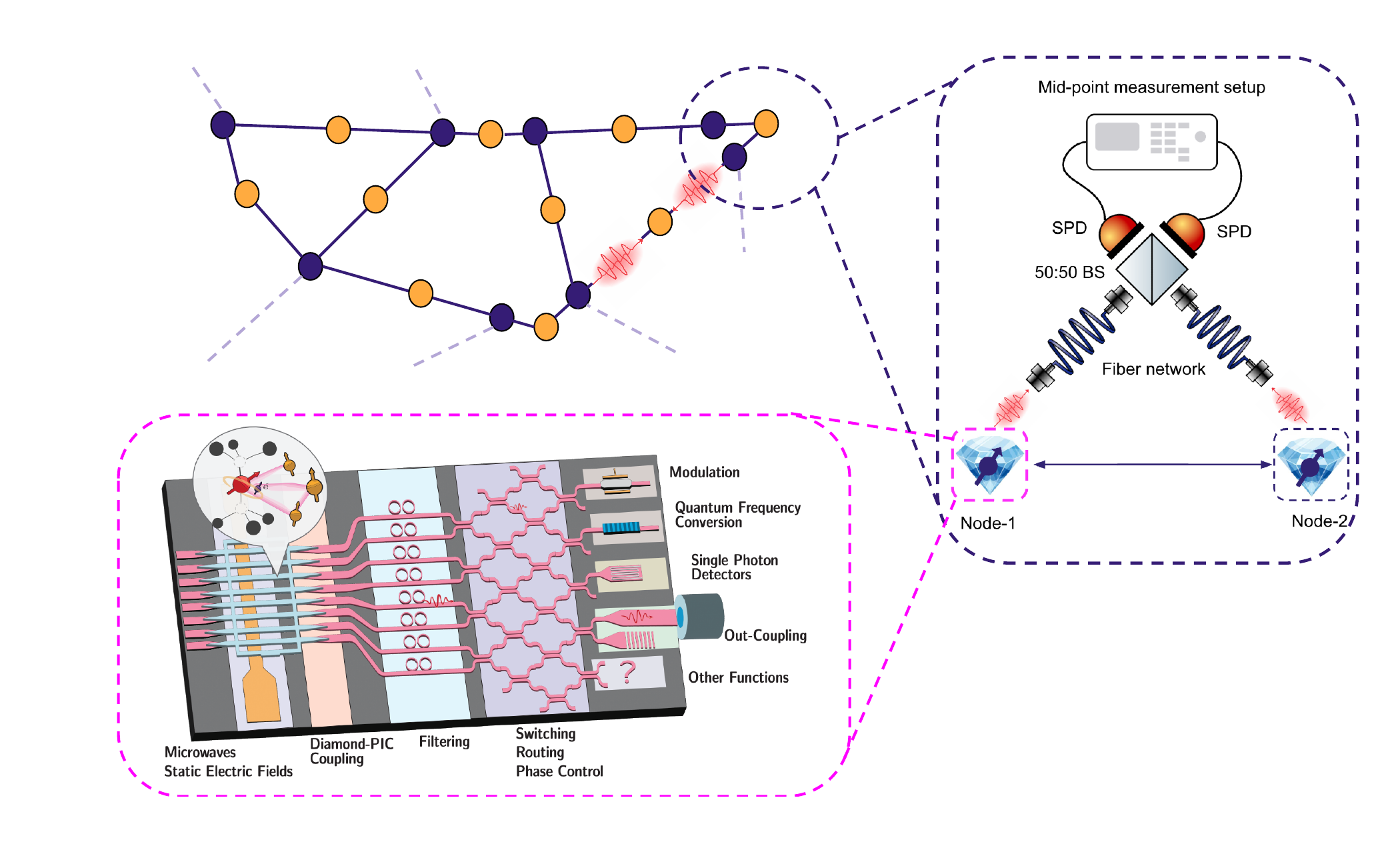}
\caption{Schematic of a large-scale quantum network composed of solid-state nodes (dark blue) that host optically interfaced communication and memory qubits. Photons are distributed through optical fibers or free-space channels to generate entanglement between distant nodes. Quantum repeaters, such as those based on midpoint-measurement–based systems, indicated by yellow circles and illustrated on the right, enhance entanglement generation rates over long distances by mitigating photon transmission losses. The bottom inset depicts a hybrid integrated quantum device functioning as a quantum node, where an optically active spin defect center with a nuclear-spin register embedded in diamond is coupled to integrated photonic and electronic circuits for optical control and spin-state manipulation of color centers. The hybrid photonic integrated circuit illustration is adapted from Ref.~\cite{pasini2024nanophotonics}.}
\label{Fig1}
\end{figure*}

The realization of large-scale quantum networks~\cite{ruf2021quantum, kimble2008quantum, bersin2024development, wehner2018quantum, chung2025interqnet} constitutes a major milestone in quantum technologies, with the potential to enable transformative applications in secure quantum communication~\cite{liao2017satellite,bhaskar2020experimental}, distributed quantum computing~\cite{nickerson2014freely, choi2019percolation}, precision sensing, and advanced metrology~\cite{gottesman2012longer, komar2014quantum}. Achieving such networks requires robust quantum nodes that combine efficient optical interfaces with long-lived and controllable qubits for information processing and storage. Figure~\ref{Fig1} illustrates a large-scale quantum network with solid-state optically active spin-qubit-based nodes.

Color defect centers~\cite{zhang2020material, atature2018material, wolfowicz2021quantum, chatterjee2021semiconductor} in diamond have emerged as leading candidates for implementing such quantum nodes~\cite{ruf2021quantum, katsumi2025recent}. Their unique combination of favorable optical properties~\cite{atature2018material, sinha2019single, eisaman2011invited} and exceptionally~long spin-coherence times~\cite{pezzagna2021quantum} enables reliable generation, storage~\cite{kalb2017entanglement, pompili2021realization}, and processing of quantum information~\cite{pezzagna2021quantum}. These characteristics make diamond defect centers particularly well-suited for scalable quantum networking architectures. Moreover, recent advances in diamond nanophotonics~\cite{katsumi2025recent, schroder2016quantum, pelucchi2022potential} have led to significant improvements in photon collection efficiency and light–matter interaction strength, paving the way toward integrated and scalable quantum processors~\cite{pezzagna2021quantum} and network nodes~\cite{ruf2021quantum, orphal2025coherent, reiserer2015cavity}.

During the past decade, substantial progress has been achieved, ranging from proof-of-principle laboratory experiments~\cite{pompili2021realization, evans2018photon, bhaskar2020experimental, bernien2013heralded} to demonstrations of entanglement distribution over metropolitan-scale fiber networks~\cite{knaut2024entanglement, stolk2024metropolitan}. Despite these remarkable advances, several key challenges remain, including limited photon indistinguishability, low entanglement generation rate, moderate entanglement fidelity, scaling-up the memory qubits~\cite{ruf2021quantum, stolk2024metropolitan, pompili2021realization, knaut2024entanglement, hermans2022qubit, iuliano2024qubit, pfaff2014unconditional}, and the integration of diamond-based quantum emitters with scalable nanophotonic~\cite{ding2024high, pregnolato2024fabrication, abulnaga2025design, chakravarthi2023hybrid, fehler2021hybrid} and photonic integrated circuit platforms~\cite{chen2024scalable, wan2020large, kim2020hybrid, palm2023modular}. Addressing these challenges is essential for the development of fault-tolerant, scalable quantum networks with high entanglement-generation rates~\cite{ruf2021quantum, nickerson2014freely, palm2023modular}.

In this review, we provide a comprehensive overview of diamond color centers in the context of quantum networking. We first discuss the requirements for the quantum network node. Then, we discuss the optical and spin properties~\cite{atature2018material, wolfowicz2021quantum, janitz2020cavity, zhang2020material, orphal2025coherent} of color centers in diamond that are relevant to network applications. We then survey recent experimental demonstrations of quantum networking at metropolitan distance scales using diamond-based platforms. Also, we identify the challenges related to the qubit systems and quantum links for metropolitan-scale networks. Finally, we discuss the promising strategies and technological advances that may enable the next generation of scalable quantum nodes for large-scale networks.

\section{An overview of quantum networks and essential requirements for a quantum network node}

An overarching objective in the development of quantum technology is to realize metropolitan-scale quantum networks capable of distributing entanglement between widely separated nodes~\cite{wehner2018quantum, kimble2008quantum, ruf2021quantum, bersin2024development}. One promising route to interconnecting stationary qubits in such large-scale networks is photon-detection-based entanglement generation~\cite{beukers2024remote}. In this distributed architecture, each node comprises quantum memories~\cite{lei2023quantum} along with quantum information processing qubits (or communication qubits) that support fast, efficient, and high-fidelity single- and two-qubit gate operations~\cite{pezzagna2021quantum}. At least one qubit per node must be optically interfaced with a shared photonic channel to enable connectivity with the other nodes~\cite{kimble2008quantum, ruf2021quantum, lei2023quantum, reiserer2015cavity}.

Through this design, quantum nodes can process, store, and transmit quantum information across the network~\cite{ruf2021quantum, reiserer2015cavity}. The availability of multiple physical qubits within a node further allows the implementation of quantum error correction~\cite{nakazato2022quantum, bartling2025universal}, leading to fault-tolerant logical qubits~\cite{bradley2019ten}. Optically active qubits can also be exploited to create high-fidelity multipartite entanglement via heralded optical measurements~\cite{wolfowicz2021quantum, lei2023quantum, atature2018material, zhang2020material}. Beyond fundamental studies, laboratory-scale quantum networks~\cite{pompili2021realization} provide a testbed to demonstrate key protocols such as nonlocal quantum gates~\cite{daiss2021quantum}, entanglement distillation~\cite{kalb2017entanglement}, and entanglement swapping~\cite{pompili2021realization}. 

\begin{figure*}[]
\centering
\includegraphics[width=\textwidth]{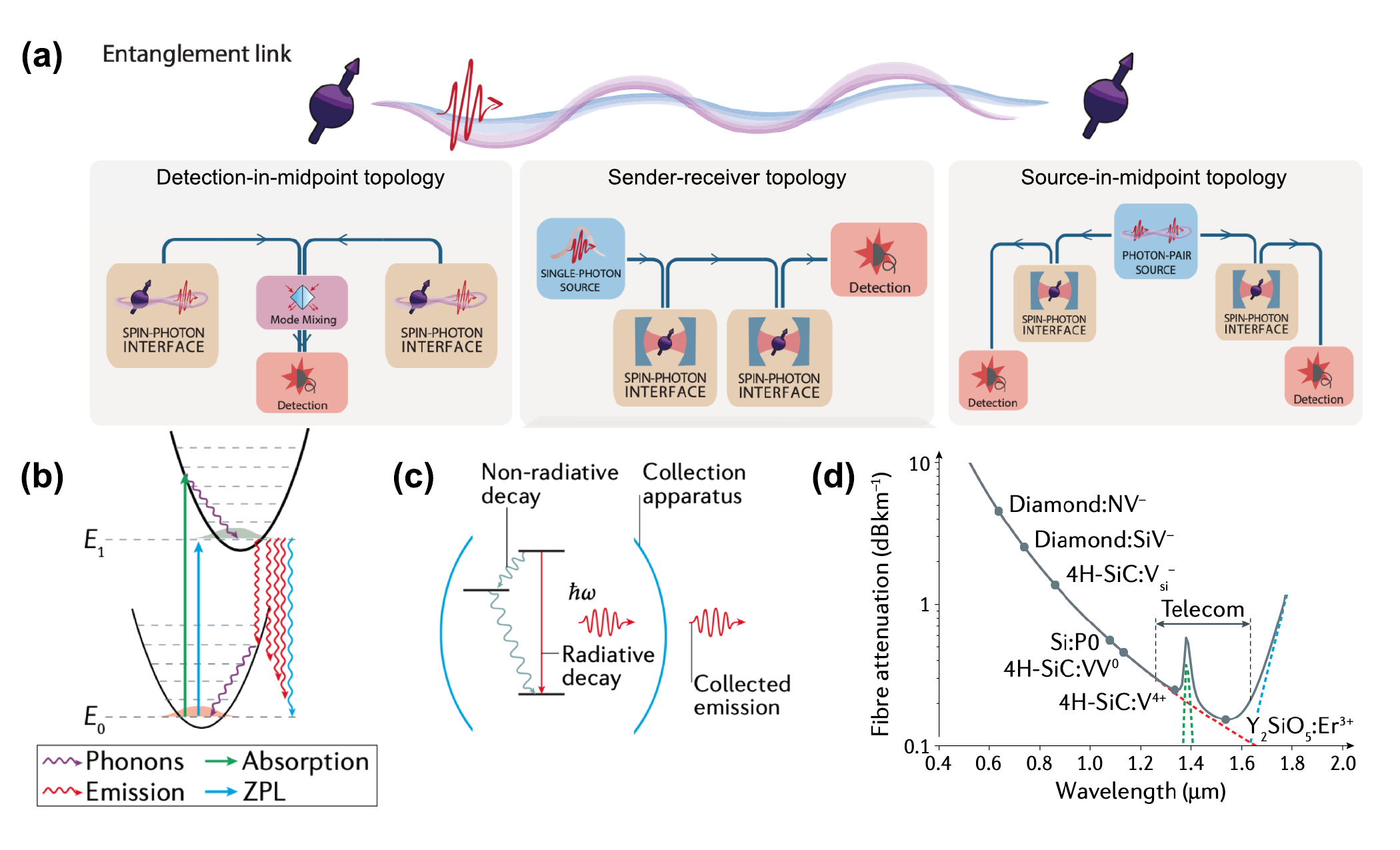}
\caption{(a) The topology for remote entanglement generation is decomposed into its fundamental logical building blocks. From left to right, the three main remote-entanglement protocol (REP) architectures are illustrated: the detection-in-midpoint, sender–receiver, and source-in-midpoint topologies. This illustration is adapted from Ref.~\cite{beukers2024remote}. (b) The potential energies of the ground and excited electronic states of an optically active solid-state quantum system are shown as functions of the collective nuclear displacement coordinate. This illustration is adapted from Ref.~\cite{wolfowicz2021quantum}. (c) Radiative and non-radiative emission of a solid-state quantum emitter in an optical collection system, i.e., photonic cavity or free-space optics. This illustration is adapted from Ref.~\cite{wolfowicz2021quantum}. (d) Attenuation loss as a function of wavelength in an optical fiber for different solid-state quantum emitters. This illustration is adapted from Ref.~\cite{wolfowicz2021quantum}.} 
\label{Fig2}
\end{figure*}

The three principal remote-entanglement protocols (REPs) for creating large-scale quantum networks, namely detection-in-midpoint, sender-receiver, and source-in-midpoint are discussed~\cite{beukers2024remote, jones2016design}. In the detection-in-midpoint scheme~\cite{stolk2024metropolitan}, both nodes generate stationary qubit–photon entanglement~\cite{togan2010quantum}, and the emitted photons are directed to a central station where they are measured in the Bell basis~\cite{atature2018material, stolk2024metropolitan}. This measurement projects the two remote stationary qubits into an entangled state~\cite{beukers2024remote}. A key advantage of this topology is its conceptual and experimental simplicity, as it fundamentally relies only on photon emission from the optically-active stationary qubit, interference at a beam splitter~\cite{waltrich2023two}, and single-photon detection at the midpoint (cf. Fig.~\ref{Fig2}(a)).

In the sender–receiver topology~\cite{beukers2024remote, jones2016design}, one node first creates stationary qubit–photon entanglement~\cite{togan2010quantum} and transmits the photon to the second node, where it interacts with another stationary qubit and photon detection establishes entanglement between two stationary qubits~\cite{knaut2024entanglement} (cf. Fig.~\ref{Fig2}(a)). This interaction can be realized through a gate-like mechanism, for example, state-dependent photon reflection of optical-cavity coupled stationary qubit system~\cite{nguyen2019integrated, chen2021polarization}. In the source-in-midpoint approach, an entangled photon-pair source \cite{sinha2019single} located at the central station, distributes photons to the two distant nodes. At each node, the reflection of a photonic qubit by an optical-cavity–coupled stationary qubit system, followed by photon measurements, projects the remote stationary qubits into an entangled state~\cite{beukers2024remote, jones2016design} (cf. Fig.~\ref{Fig2}(a)).

To realize a suitable material platform~\cite{zhang2020material, atature2018material, wolfowicz2021quantum, chatterjee2021semiconductor} for quantum nodes in large-scale quantum networks, stationary qubits serving as information-processing and memory qubits~\cite{ruf2021quantum}. The memory qubits must be capable of storing quantum states during the entanglement generation process, which requires coherence times under full network operation to exceed the entanglement generation time. Also, each node should support the storage of multiple entangled states and allow high-fidelity operations between them, enabling multiqubit protocols such as quantum error correction~\cite{wolfowicz2021quantum, lei2023quantum, atature2018material, zhang2020material}.

In Table~\ref{Table-1.1}, the physical characteristics of various solid-state quantum systems are summarized and compared from the perspective of their suitability as quantum nodes~\cite{wolfowicz2021quantum, lei2023quantum, atature2018material, zhang2020material, janitz2020cavity, bhunia2024ultrafast}. Key node-level performance metrics, such as storage time, processor scalability, entanglement generation rate, state transfer rate, and quantum gate fidelity, are strongly influenced by several underlying material, spin and optical parameters~\cite{atature2018material, lei2023quantum}. These include the number and interaction properties of the ancilla qubits, operating temperature, the coherence time ($T_2$) of both communication and memory qubits, the zero-phonon-line (ZPL) wavelength, Debye-Waller factor, linewidth of optical transition, excited-state lifetime, and quantum efficiency~\cite{wolfowicz2021quantum, lei2023quantum, atature2018material, zhang2020material, janitz2020cavity}.

The operating temperature is defined as the temperature below which the expected orbital lifetime exceeds $100~\mathrm{ms}$~\cite{ruf2021quantum, beukers2025quantum}. This temperature can be estimated from the phonon transition rate

\begin{equation}
\frac{\Gamma_+}{\Delta_{gs}^3} \propto \frac{1}{e^{\Delta_{gs}/k_B T} - 1}\ ,
\end{equation}

where $k_B$ is the Boltzmann constant and $\Gamma_+$ determines the coherence time of the qubit. Consequently, operating temperature is a crucial parameter for maintaining the coherence of the communication-qubit, while the achievable storage time is directly determined by the coherence time ($T_2$) of the memory qubits~\cite{ruf2021quantum, beukers2025quantum}.

Furthermore, the Debye-Waller factor~\cite{wolfowicz2021quantum, atature2018material, zhang2020material} and the linewidth of the ZPL transition govern the number of indistinguishable photons~\cite{wolfowicz2021quantum, atature2018material, zhang2020material} and their spectral indistinguishability in ZPL emission~\cite{wolfowicz2021quantum, atature2018material, zhang2020material}, respectively (cf.\ Fig.~\ref{Fig2}(b)). The optical coherence of a quantum emitter is characterized by the coherence time $\tau_2$, governed by the excited-state lifetime $\tau_1$ and pure dephasing $\tau_2^*$. The lifetime $\tau_1$ sets a fundamental upper bound $\tau_2 \leq 2\tau_1$ (the transform limit), where the optical linewidth reaches its minimum $\Delta\nu = 1/2\pi\tau_1$ \cite{wolfowicz2021quantum, esmann2024solid}. Quantum efficiency, defined as the ratio of radiative decay to all decay channels, also plays an important role in spin-photon interface \cite{radko2016determining, ruf2021quantum}. Collectively, these parameters critically affect the entanglement generation rate, state transfer efficiency, and overall fidelity of quantum networking protocols~\cite{wolfowicz2021quantum, atature2018material, zhang2020material}.

\begin{table*}[htbp]
\centering
\caption{\bf Comparison of key physical properties of representative solid-state quantum platforms relevant to quantum-node applications. Here, e-spin and n-spin denote electronic spin and nuclear spin, respectively.}

\begin{tabular}{cccccc}
\hline
System &
Ancilla qubits &
Operating &
Spin &
Optical transition\\

 &
 &
temperature (K) &
coherence time ($T_2$) &
wavelength (nm) \\

\hline

NV-center (diamond) &
$^{13}$C, $^{14}$N \cite{pezzagna2021quantum, bradley2019ten} &
- &
$1.5$ s (e-spin) \cite{abobeih2018one} &
$637$ (ZPL) \cite{doherty2013nitrogen, santori2006coherent}\\

&
&
&
$1$ min (n-spin) \cite{bradley2019ten} &
\\

SiV-center (diamond) &
$^{29}$Si, $^{13}$C \cite{knaut2024entanglement} &
$0.1$  \cite{ruf2021cavity} &
$13$ ms (e-spin) \cite{sukachev2017silicon} &
$737$ (ZPL) \cite{hepp2014electronic, ruf2021quantum}\\

&
&
&
$2$ s (n-spin) \cite{stas2022robust} &
\\

GeV-center (diamond) &
$^{29}$Ge, $^{13}$C \cite{atature2018material}&
$0.4$ \cite{ruf2021quantum} &
$24$ ms (e-spin) \cite{orphal2025coherent, senkalla2024germanium}&
$602$ (ZPL) \cite{orphal2025coherent, ruf2021quantum}\\

&
&
&
$2.5$ s (n-spin) \cite{grimm2025coherent} &
\\

SnV-center (diamond) &
$^{118}$Sn, $^{119}$Sn \cite{narita2023multiple}&
$1.8$ \cite{ruf2021quantum}&
$10$ ms (e-spin) \cite{orphal2025coherent, metsch2019initialization, karapatzakis2024microwave} &
$620$ (ZPL) \cite{ruf2021quantum, orphal2025coherent}\\

 &
 $^{13}$C \cite{karapatzakis2024microwave} &
 &
 $1.35$ s (n-spin) \cite{resch2026high} &
\\

SiC Divacancy &
$^{29}$Si, $^{13}$C \cite{atature2018material}&
- &
$\sim 1$ ms \cite{atature2018material}, $64$ ms \cite{lukin2020integrated}&
$1100$ (ZPL) \cite{atature2018material}\\

$V_{\mathrm{Si}}$ (SiC) &
$^{29}$Si, $^{13}$C \cite{atature2018material}&
- &
$\sim 20$ ms \cite{atature2018material}&
$900$ (ZPL) \cite{atature2018material} \\

Quantum dots (TMDCs) &
- &
- &
- &
$1080-1550$ \cite{zhao2021site, atature2018material}\\

InAs/GaAs quantum dots &
Mn dopants &
- &
$3-10$ $\mu$s \cite{atature2018material, somaschi2016near}&
$900-1550$ \cite{atature2018material, somaschi2016near}\\

 &
 coupled QDs \cite{atature2018material}&
 &
 & \\

RE crystal ($^{167}$Er$^{3+}$:Y$_2$SiO$_5$ \cite{ranvcic2018coherence})&
- &
$<4$ \cite{ranvcic2018coherence, lei2023quantum, saglamyurek2015quantum}&
$1.3$ s (n-spin) \cite{ranvcic2018coherence}&
$1532$ \cite{lei2023quantum, saglamyurek2015quantum},\\

&
&
&
&
$1536.14$, $1538.57$ \cite{gritsch2022narrow}\\

RE crystal (Pr$^{3+}$:Y$_2$SiO$_5$ \cite{heinze2013stopped})&
- &
$<4$ \cite{heinze2013stopped, lei2023quantum, duranti2024efficient}&
$30$ s (n-spin) \cite{heinze2013stopped}&
$606$ \cite{heinze2013stopped, lei2023quantum, duranti2024efficient} \\

RE crystal (Eu$^{3+}$:Y$_2$SiO$_5$ \cite{ma2021one, wang2025nuclear})&
- &
$<4$ \cite{ma2021one, wang2025nuclear, lei2023quantum, ortu2022storage}&
$\sim 6$ h (n-spin) \cite{wang2025nuclear}&
$580$ \cite{ma2021one, wang2025nuclear, lei2023quantum, ortu2022storage}\\

RE crystal ($^{171}$Yb$^{3+}$:YVO \cite{kindem2020control})&
- &
$<4$ \cite{kindem2020control}&
$30$ ms (n-spin) \cite{kindem2020control}&
$984.5$ \cite{kindem2020control}\\

\hline
\end{tabular}
  \label{Table-1.1}
  
$^\textit{}$
\end{table*}
\begin{table*}[htbp]
\centering
\begin{tabular}{ccccccc}
\hline
System &
Debye-Waller factor &
Linewidth &
Lifetime (ns) &
Quantum efficiency \\
\hline
NV-center (diamond) &
$\sim 0.03$ \cite{doherty2013nitrogen, santori2006coherent}&
$13$ MHz\cite{doherty2013nitrogen, santori2006coherent}&
$12$ \cite{doherty2013nitrogen, santori2006coherent}&
$>0.8$ \cite{ruf2021quantum}\\

SiV-center (diamond) &
$0.65-0.9$ \cite{hepp2014electronic, ruf2021quantum}&
$\sim 100$ MHz\cite{wang2006single, hepp2014electronic}&
$1.2-1.7$ \cite{wang2006single, hepp2014electronic}&
$0.01-0.1$ \cite{ruf2021cavity, sipahigil2016integrated}\\

GeV-center (diamond) &
$\sim 0.6$ \cite{siyushev2017optical, ruf2021quantum}&
$\sim 45$ MHz\cite{orphal2025coherent, senkalla2024germanium}&
$1.1-6$ \cite{orphal2025coherent, senkalla2024germanium}&
$0.12-0.25$ \cite{hoy2020cavity, ruf2021quantum}\\

SnV-center (diamond) &
$\sim 0.57$ \cite{ruf2021quantum, orphal2025coherent}&
$27$ MHz\cite{orphal2025coherent}&
$4.5-7$ \cite{orphal2025coherent}&
$\sim 0.8$ \cite{ruf2021quantum, orphal2025coherent}\\

SiC Divacancy &
$\sim 0.07$ \cite{atature2018material}&
$>80$ MHz\cite{atature2018material}&
$90$ \cite{lukin2020integrated}&
- \\

$V_{\mathrm{Si}}$ (SiC) &
$0.06–0.09$ \cite{lukin2020integrated, atature2018material}&
$51$ MHz \cite{lukin2020integrated}&
$37$ \cite{lukin2020integrated}&
- \\

Quantum dots (TMDCs) &
- &
$\sim 2$ GHz\cite{atature2018material}&
$5$ ns\cite{atature2018material}&
- \\

InAs/GaAs quantum dots &
$>0.95$ \cite{atature2018material, somaschi2016near}&
$\sim 300$ MHz \cite{atature2018material, somaschi2016near}&
$4$ ns\cite{atature2018material}&
$\sim 1$ \cite{atature2018material, somaschi2016near}\\

RE crystal ($^{167}$Er$^{3+}$:Y$_2$SiO$_5$ \cite{ranvcic2018coherence}) &
- &
$0.2$ GHz \cite{holewa2025solid} &
$1.14\times10^{7}$ \cite{holewa2025solid} &
- \\

RE crystal (Pr$^{3+}$:Y$_2$SiO$_5$ \cite{heinze2013stopped})&
- &
$2.8$ kHz \cite{equall1995homogeneous}&
$164\times 10^3$ \cite{equall1995homogeneous}&
- \\

RE crystal (Eu$^{3+}$:Y$_2$SiO$_5$ \cite{ma2021one, wang2025nuclear})&
- &
sub-GHz to $150$ GHz \cite{konz2003temperature}&
$1.97\times 10^{6}$ \cite{konz2003temperature}&
- \\

RE crystal ($^{171}$Yb$^{3+}$:YVO \cite{kindem2020control})&
- &
$31$ GHz\cite{kindem2020control}&
$267\mu$s \cite{kindem2020control}&
- \\

\hline
\end{tabular}
  
$^\textit{}$
\end{table*}

A color center hosts an individually addressable, optically active electronic spin that serves as a communication qubit for quantum information processing, while also providing access to multiple long-lived nearby nuclear spins that function as memory qubits~\cite{pezzagna2021quantum}. These nuclear spins can be controlled with high fidelity, allowing the communication qubit to be released for further operations~\cite{knaut2024entanglement}, allowing the implementation of multiqubit protocols. Their long coherence times~\cite{atature2018material} support reliable quantum state storage during ongoing network operations \cite{yang2016high}.

The intrinsic energy-level structure of the color centers~\cite{beukers2024remote, maze2011properties, pasini2024nanophotonics} is well suited for remote entanglement generation. In particular, color centers exhibit spin-state-selective optical transitions~\cite{atature2018material, maze2011properties, togan2010quantum} that enable entanglement between the electronic spin and a photonic degree of freedom, such as photon number, polarization~\cite{togan2010quantum}, or time-bin encoding~\cite{bhaskar2020experimental}. Color centers also exhibit bright optical emission that can be collected efficiently (cf.\ Fig.~\ref{Fig2}(c)), leading to high entanglement generation rates. This emission typically lies in the visible to near-infrared (NIR) spectral range, where the optical fiber losses are higher than in the telecommunications band (cf.\ Fig.~\ref{Fig2}(d)). Nevertheless, efficient frequency conversion to telecommunications wavelengths can be achieved while preserving quantum correlations~\cite{ruf2021quantum, bersin2024telecom, iuliano2024qubit, schafer2025two, dreau2018quantum}.

Based on the requirements and material properties discussed above, diamond defect centers have emerged as promising candidates for quantum-node applications~\cite{wolfowicz2021quantum, lei2023quantum, atature2018material, zhang2020material, janitz2020cavity}. In recent years, significant progress has been made in diamond nanophotonics~\cite{wan2020large, kim2020hybrid, chen2024scalable}. These devices can be integrated with photonic circuits (cf. Fig.~\ref{Fig1}), and efficient optical interfacing with embedded color centers in such architectures has already been demonstrated~\cite{wan2020large, chen2024scalable, kim2020hybrid}.

\section{Optical and spin properties of color defects in diamond}

Quantum nodes can act as quantum computers for processing quantum information~\cite{pezzagna2021quantum, ruf2021quantum}. An efficient quantum node must be strongly isolated from its environment \cite{steger2012quantum, harris2024coherence, wolfowicz2021quantum}, while still permitting precise external control~\cite{orphal2025coherent}. Perfect isolation and perfect controllability cannot be achieved simultaneously; instead, practical implementations rely on a trade-off that modestly limits the fidelity of both. An effective way to balance these requirements is to place the quantum system in a vacuum, where environmental perturbations are minimized, while qubits can still be coherently manipulated using electromagnetic fields~\cite{pezzagna2021quantum}.

Solid-state host materials~\cite{zhang2020material, atature2018material, chatterjee2021semiconductor} represent a particularly promising platform for quantum-node applications, especially due to their compatibility with modern CMOS electronics~\cite{kim2020hybrid, pelucchi2022potential, li2024heterogeneous}. Among all solid-state material platforms (cf. Table~\ref{Table-1.1}), the high-purity, homoepitaxially chemical-vapour-deposition (CVD) grown diamond (type-IIa) with a natural abundance of carbon isotopes \((1.1\% \ ^{13}C)\) stands out \cite{bradley2019ten}. Due to its wide band gap of \(5.4\)~eV, the diamond exhibits vacuum-like properties, such that a point defect center can be viewed as an atom confined in a trap~\cite{pezzagna2021quantum, ruf2021quantum, atature2018material, zhang2020material}. The large band gap ensures that the conduction band remains unoccupied even at room temperature, preventing unwanted interactions between free carriers and the qubits. Strong covalent bonding leads to a high Debye temperature, exceeding $1800$ K, significantly suppressing phonon-induced decoherence ~\cite{doherty2012theory, doherty2013nitrogen, wolfowicz2021quantum}. Moreover, the low diffusion constants of most elements allow atomic positions to remain stable even at temperatures of several hundred kelvin~\cite{pezzagna2021quantum, ruf2021quantum, atature2018material, zhang2020material}.

Because defect-based qubits are defined by single atoms with well-controlled structures, quantum devices fabricated in this manner exhibit highly reproducible electronic structures, optical transition frequencies, and spin properties~\cite{chatterjee2021semiconductor, wolfowicz2021quantum, de2021materials, awschalom2025challenges}. This uniformity is a key advantage over solid-state quantum devices based on quantum dots, where variations in size, composition, and atomic configuration lead to device-to-device inhomogeneity~\cite{pezzagna2021quantum, ruf2021quantum, atature2018material, zhang2020material}.

Here, the optical and spin properties of the negatively charged nitrogen-vacancy (NV) center (cf. Fig.~\ref{Fig3}(a)), along with other negatively charged group-IV defect centers in diamond (cf. Fig.~\ref{Fig4}(a)), i.e., silicon (Si) vacancy, germanium (Ge) vacancy, and tin (Sn) vacancy, are discussed from a quantum networking perspective~\cite{ruf2021quantum, atature2018material, orphal2025coherent}.

\subsection{Optical and spin properties of NV center in diamond}

\begin{figure*}[]
\centering
\includegraphics[width=\textwidth]{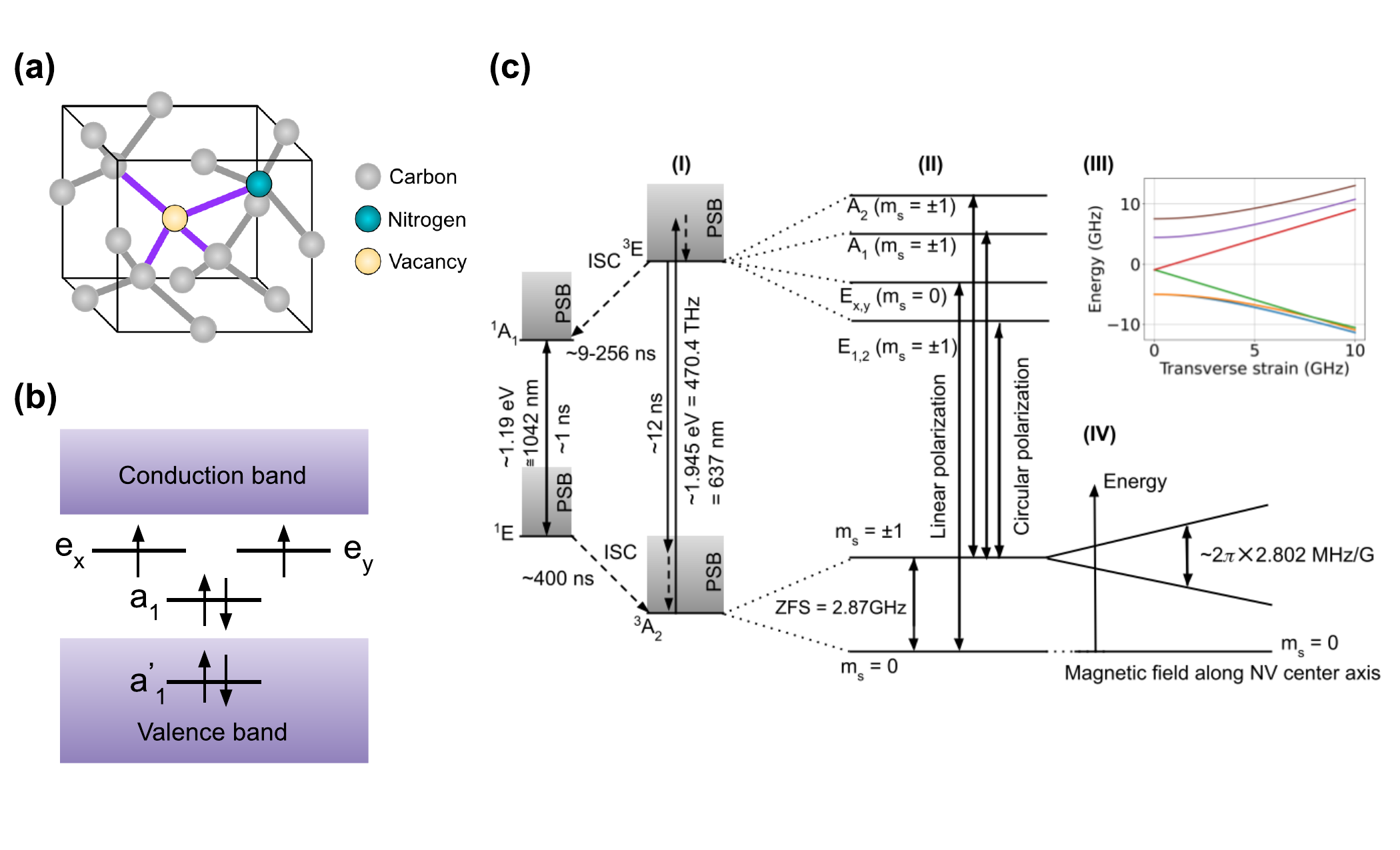}
\caption{(a) Atomic configuration of the NV defect inside a diamond lattice unit cell. (b) Molecular orbitals (MO) and corresponding electronic distribution in the ground state of the NV center \cite{ruf2021cavity}. (c) Electronic energy-level diagram of the negatively charged nitrogen vacancy defect center in a diamond crystal. This illustration is adapted from Refs.~\cite{beukers2025quantum, pasini2024nanophotonics}.}
\label{Fig3}
\end{figure*}

Figure~\ref{Fig3}(b) shows the ground-state molecular orbitals (MOs) of the NV center that arise from linear combinations of the atomic orbitals (LCAOs), which are consistent with the $C_{3v}$ symmetry of the defect~\cite{doherty2012theory, doherty2013nitrogen}. For the NV center in its ground state, the MOs are occupied as ${a'_1}^2 a_1^2 e^2$, whereas in the excited state, an electron from the $a_1$ orbital is transferred to the degenerate $e$ orbitals, thus yielding the configuration ${a'_1}^2 a_1 e^3$. In both states, two unpaired electrons reside within the diamond band gap, giving rise to atom-like coherence properties despite the solid-state environment \cite{atature2018material, ruf2021quantum, chatterjee2021semiconductor}.

Coulomb repulsion lifts the degeneracy of the spin singlet and triplet manifolds, producing the electronic level structure illustrated in Fig.~\ref{Fig3}(c)-(I)~\cite{ruf2021cavity, atature2018material, ruf2021quantum, maze2011properties}. Optical transitions can occur directly through the ZPL or indirectly through vibrational excitations that give rise to the phonon sideband (PSB)~\cite{ruf2021cavity}. This behavior can be described by the Huang-Rhys model \cite{alkauskas2014first}, which considers electron-phonon coupling and its effect on transition probabilities~\cite{atature2018material, ruf2021quantum}.

Because the ground and excited states exhibit different charge distributions, their equilibrium nuclear positions and vibrational spacings differ slightly~\cite{ruf2021cavity}. Under the Born-Oppenheimer approximation, optical transitions occur much faster than nuclear motion, so transitions often involve states with different vibrational occupations~\cite{zhang2020material, wolfowicz2021quantum}. The Franck-Condon principle dictates that the probability of a transition depends on the overlap of vibrational wavefunctions. As a result, the NV emission spectrum features a sharp ZPL and several red-shifted PSBs (cf. Fig.~\ref{Fig5}(a)). Only about $2.6\%$ of the emission is contained in the ZPL with the majority involving phonon-assisted processes~ \cite{zhang2020material, ruf2021cavity, atature2018material}.

At room temperature, the ZPL linewidth broadens due to mixing of the excited states through phonon interactions. A $T^5$-dependent two-phonon Raman-scattering process~\cite{doherty2013nitrogen, fu2009observation} governs these interactions and broadens the PSB transitions due to the short lifetimes of lattice vibrations. At cryogenic temperatures, the fine structure of the excited state (cf. Fig.~\ref{Fig5}(a)) becomes resolvable, enabling resonant, spin-selective optical addressing.

Transitions between ground and excited states follow polarization-dependent selection rules and are spin-conserving~\cite{togan2010quantum, beukers2025quantum, maze2011properties}. These properties enable the generation of spin-photon entanglement~\cite{togan2010quantum}. Table~\ref{Table-2} summarizes the selection rules for transitions between the triplet states.

\begin{table}[htbp]
\centering
\caption{\bf Selection rules for optical transitions between the orbital-doublet spin-triplet excited states (\(A_1, A_2, E_1, E_2, E_x, E_y\)) and the orbital-singlet spin-triplet ground states (\(^3A_{2-}, ^3A_{20}, ^3A_{2+}\)). Circular polarizations are denoted by \(\hat{\sigma}^{\pm} = \hat{x} \pm i\hat{y}\), and linear polarizations by \(\hat{x}, \hat{y}\) \cite{togan2010quantum, maze2011properties}.}
\begin{tabular}{ccccccc}
\hline
Electronic spin-states & $A_1$ & $A_2$ & $E_1$ & $E_2$ & $E_x$ & $E_y$ \\
\hline
$^3A_{2-}$  
& $\hat{\sigma}^+$ & $\hat{\sigma}^+$ 
& $\hat{\sigma}^-$ & $\hat{\sigma}^-$ 
& {-} & {-} \\
$^3A_{20}$  
& {-} & {-} & {-} & {-} 
& $\hat{y}$ & $\hat{x}$ \\
$^3A_{2+}$  
& $\hat{\sigma}^-$ & $\hat{\sigma}^-$ 
& $\hat{\sigma}^+$ & $\hat{\sigma}^+$ 
& {-} & {-} \\
\hline
\end{tabular}
  \label{Table-2}
  $^\textit{}$
\end{table}

Consideration of spin–spin and spin-orbit interactions results in the energy-level diagram of the NV center, which is shown in Fig.~\ref{Fig3}(c)-II. Even without an external magnetic field, the orbital-singlet spin-triplet ground state exhibits an intrinsic energy-level splitting. Whereas electron spins are aligned ($m_s = \pm 1$), a zero-field splitting (ZFS) of $D \approx 2.87~\text{GHz}$ occurs relative to the anti-parallel spin configuration ($m_s = 0$) due to the spin–spin interaction. Applying a magnetic field along the NV axis lifts the degeneracy of the $m_s = \pm 1$ states through the Zeeman effect, defining a qubit subspace (cf. Fig.~\ref{Fig3}(c)-(IV)). Under cryogenic conditions and with the implementation of dynamical decoupling (DD) techniques, coherence times  $T_2 > 1~\text{s}$ have been demonstrated~\cite{ruf2021quantum, atature2018material, orphal2025coherent}.

The nitrogen nucleus in the NV center exhibits a spin of $I = 1$ for $^{14}\mathrm{N}$ (concentration \(\approx 99.3\%\)) or $I = \tfrac{1}{2}$ for $^{15}\mathrm{N}$ (concentration \(\approx 0.7\%\)), resulting in an extra ground-state splitting on the order of $2~\text{MHz}$ \cite{ruf2021quantum, bradley2019ten}. When broadband microwave (MW) pulses with sufficiently high Rabi frequencies are applied, all nuclear spin sublevels can be driven simultaneously. Alternatively, the nitrogen nuclear spin can serve as a separate qubit, controlled by specific radio-frequency (RF) pulses~\cite{orphal2025coherent, bartling2025universal}.

The NV center electron spin also interacts with nearby $^{13}$C nuclear spins ($I = 1/2$, $1.1\%$ natural abundance) via hyperfine interactions~\cite{bradley2019ten, bartling2025universal}. These $^{13}$C spins act as long-lived memory qubits with coherence times exceeding $10~\text{s}$~\cite{pezzagna2021quantum, bradley2019ten} and can be coherently controlled using DD sequences or RF pulses~\cite{orphal2025coherent, bradley2021order}.

The NV center electron spin also interacts with nearby $^{13}$C nuclear spins ($I = 1/2$, $1.1\%$ natural abundance) via hyperfine interactions~\cite{bradley2019ten, bartling2025universal}. These $^{13}$C spins act as long-lived memory qubits with coherence times exceeding $10~\text{s}$~\cite{pezzagna2021quantum, bradley2019ten} and can be coherently controlled using DD sequences or RF pulses~\cite{orphal2025coherent, bradley2021order}.

In the excited electronic state accessed optically, the orbital-doublet spin-triplet manifold is divided into six levels by the combined effects of spin–spin and spin–orbit coupling. Transverse strain or electric fields can lift degeneracies by distorting the orbital wavefunctions (cf. Fig.~\ref{Fig3}(c)-III), while axial strain instead shifts the entire manifold relative to the ground state~\cite{maze2011properties, beukers2025quantum}.

Spin-state-dependent decay through intermediate singlet states enables efficient optical spin polarization even at room temperature (300 K)~\cite{orphal2025coherent, suter2017single}. This same mechanism forms the basis for optically detected magnetic resonance (ODMR), in which NV centers in the $m_s = \pm 1$ states emit fewer photons than those in the $m_s = 0$ state, owing to the differing intersystem crossing (ISC) probabilities associated with each spin state.

The NV center forms a composite spin system consisting of the electronic spin, the intrinsic $^{14}$N or $^{15}$N nuclear spin, and a surrounding bath of many $^{13}$C nuclear spins~\cite{bradley2021order, bradley2019ten}. In the electronic ground state, the total Hamiltonian can be written as the sum of the individual spin Hamiltonians (electron \cite{suter2017single, ruf2021cavity, orphal2025coherent, rondin2014magnetometry}, $^{14}$N or $^{15}$N \cite{bradley2021order, degen2021creation}, and $^{13}$C \cite{bradley2021order}) and their mutual interactions like, electron-nuclear spin interaction (hyperfine interaction) and dipolar interaction among nuclear spins \cite{bradley2021order, bradley2019ten}. 

A combination of MW and RF pulses can be implemented to control the electron and nuclear spins of the NV center~\cite{bradley2021order, orphal2025coherent, bradley2019ten}. The electron spin exhibits an exceptional spin-lattice relaxation time that exceeds 1\,h at 4\,K~\cite{abobeih2018one, pezzagna2021quantum}. Furthermore, by suppressing unwanted interactions with the crystal environment using tailored MW pulse sequences, quantum information can be stored in the electronic spin qubit for more than 1.5\,s at 4\,K~\cite{abobeih2018one, pezzagna2021quantum}. In addition, the $^{13}$C nuclear spin serves as an excellent memory qubit, owing to its remarkable spin-lattice relaxation (\(T_1\)) and coherence times (\(T_2\)) of approximately 6\,min~\cite{pezzagna2021quantum, bradley2019ten} and 1\,min~\cite{pezzagna2021quantum, bradley2019ten}, respectively, at 4\,K. 
Beyond the coherence properties, the NV center platform also demonstrates outstanding gate performance with single-qubit and two-qubit gate fidelities reaching about 99.995$\%$~\cite{pezzagna2021quantum, rong2015experimental} and exceeding 97$\%$~\cite{yamamoto2013strongly, pezzagna2021quantum}, respectively.

\subsection{Optical and spin properties of negatively-charged group-IV color centers  in diamond}

\begin{figure*}[]
\centering
\includegraphics[width=\textwidth]{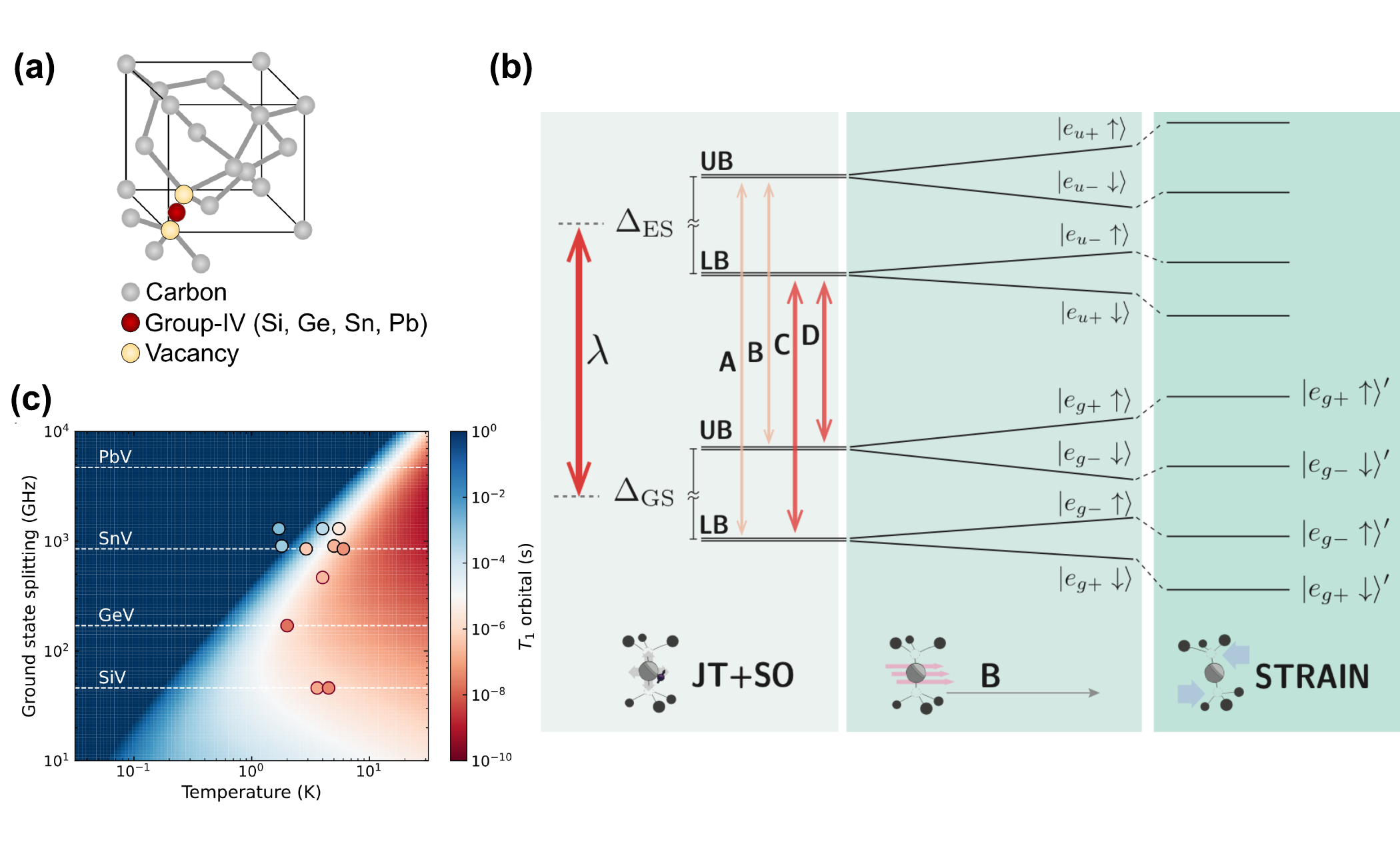}
\caption{(a) Atomic configuration of a negatively charged group-IV defect within a diamond lattice unit cell. (b) The orbital coherence time $T_1^{\rm orbital}$ depends on the ground state splitting and temperature. The colormap gives the simulated values using $\chi \rho = 4.6 \times 10^{-9}$\,GHz$^{-2}$. The circles indicate measured values of SiV, GeV, and SnV centers. The $T_1^{\rm spin}$ and $T_2^{\rm *,spin}$ values have been converted to the corresponding $T_1^{\rm orbital}$. The experimental values can be lower than the simulations as the coherence can be limited by other sources. This illustration is adapted from Ref.~\cite{beukers2025quantum}. (c) The electronic level stucture of negatively charged group-IV defect centers in diamond. This figure is adapted from Ref.~\cite{beukers2025quantum}.}
\label{Fig4}
\end{figure*}

\begin{figure*}[]
\centering
\includegraphics[width=0.48\linewidth]{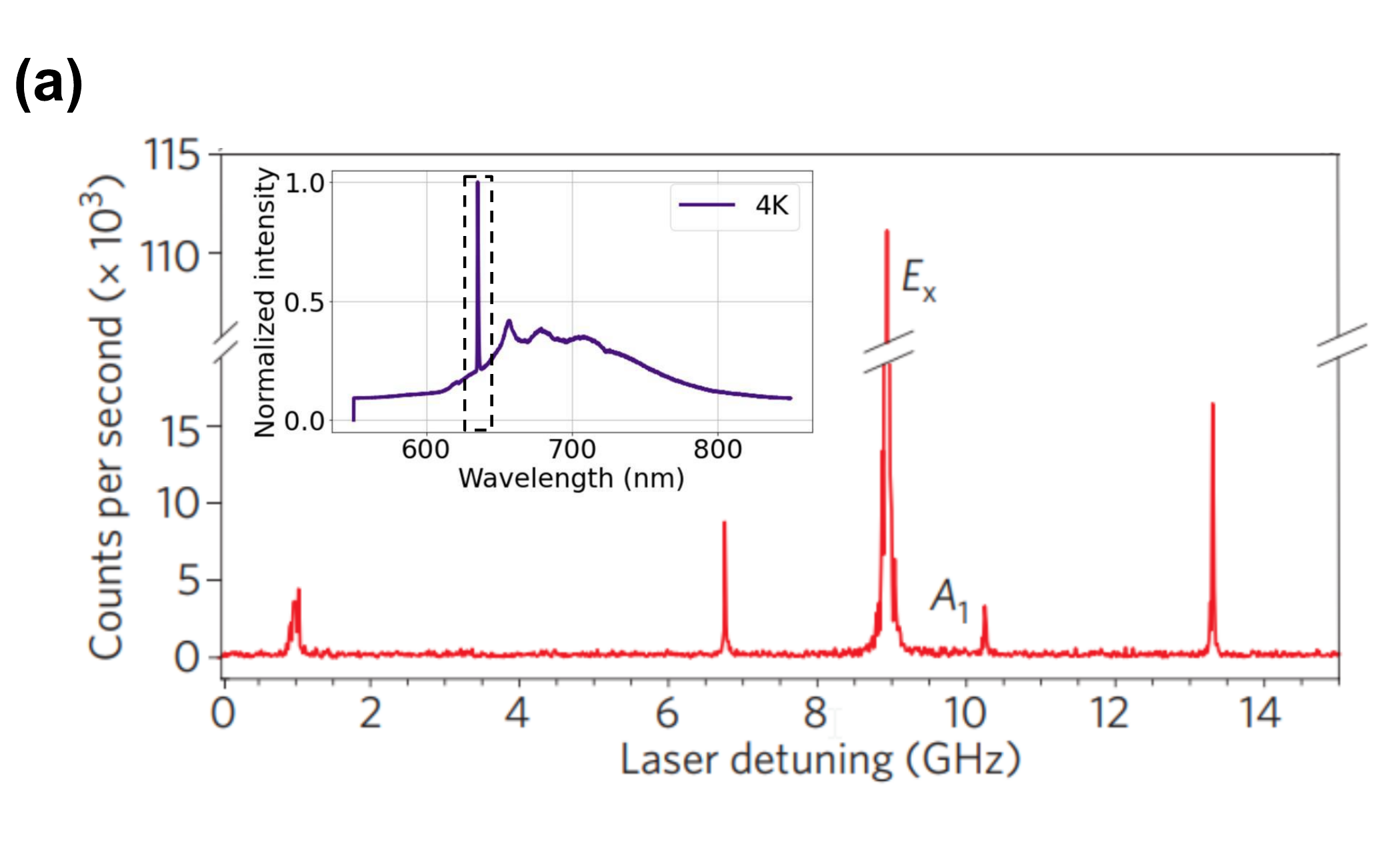}
\includegraphics[width=0.48\linewidth]{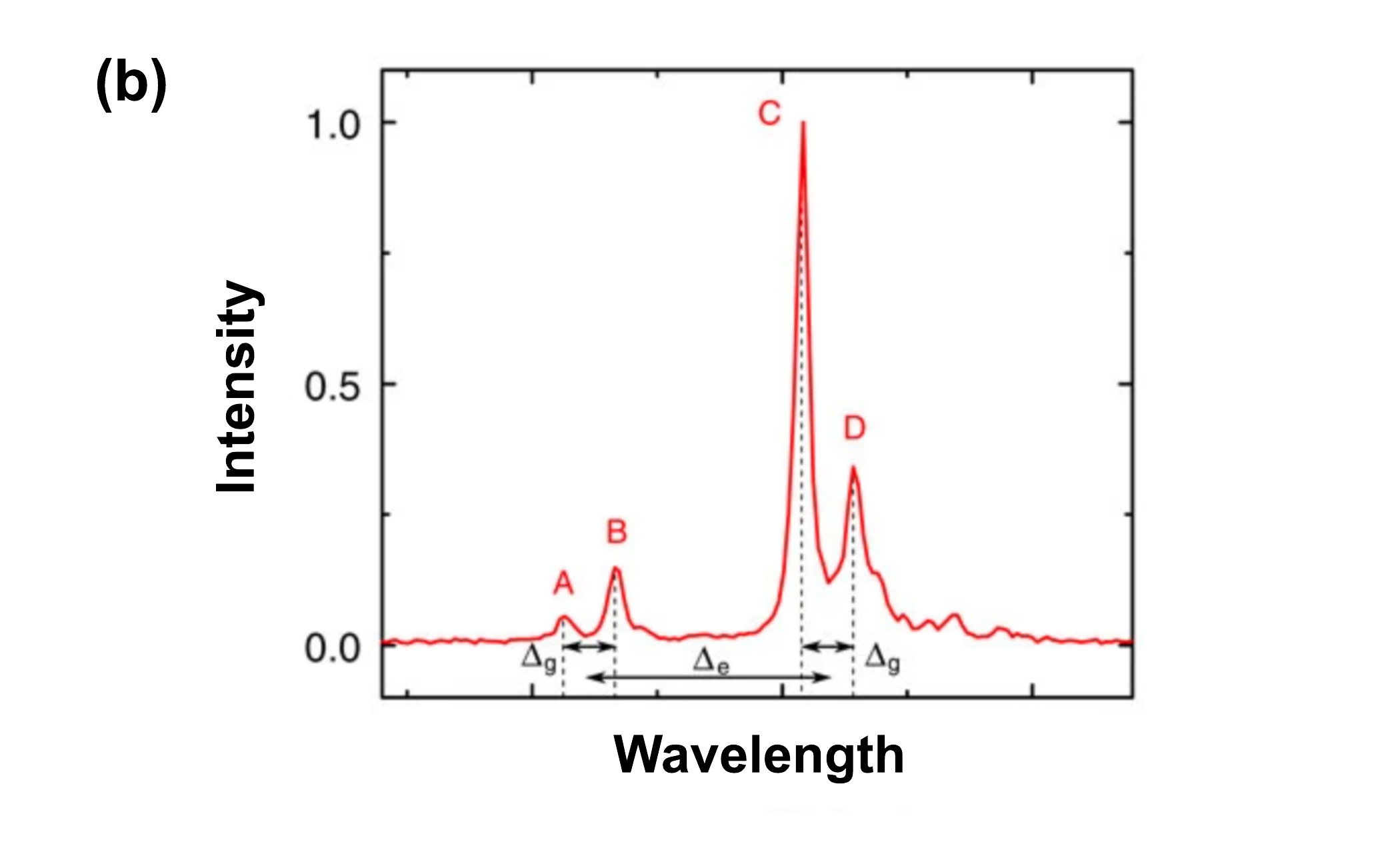}
\caption{(a) Excited-state transitions of the NV center were resolved using a tunable red laser scanned across the resonance. This illustration is adapted from Ref.~\cite{bogdanovic2017diamond}. The inset figure shows the emission spectrum of the NV centers at 4\,K under 532\,nm laser excitation. (b) Low-temperature emission spectra of group-IV color centers in diamond. This figure is adapted from Ref.~\cite{muller2014optical}.}
\label{Fig5}
\end{figure*}

Despite the outstanding spin properties, the fidelity of quantum networking protocols based on the NV center is limited by its optical characteristics, primarily the low Debye-Waller factor of about 3$\%$ (cf. Fig.~\ref{Fig5}(a)) and fluctuations of the ZPL arising from local strain or nearby charges~\cite{ruf2021quantum, atature2018material, janitz2020cavity}. An alternative family of color defects in diamond is provided by the group-IV vacancy centers~\cite{ruf2021cavity, atature2018material, janitz2020cavity, zhang2020material}. These centers are formed when a group-IV atom, such as Si~\cite{hepp2014electronic, ruf2021quantum}, Ge~\cite{iwasaki2015germanium}, Sn ~\cite{iwasaki2017tin, trusheim2020transform}, or lead (Pb) ~\cite{trusheim2019lead, ditalia2018single}, occupies an interstitial position between two adjacent vacancies in the diamond lattice (cf. Fig.~\ref{Fig4}(a)). The resulting inversion-symmetric configuration with $D_{3d}$ point-group symmetry suppresses the formation of a permanent electric dipole moment and consequently inhibits first-order DC Stark shifts~\cite{orphal2025coherent, de2021investigation, aghaeimeibodi2021electrical}. Due to this reason, the group-IV vacancy centers show outstanding optical coherence properties and more than $50\%$~\cite{ruf2021quantum, atature2018material, orphal2025coherent} of the emission spectrum consists of coherent photons in the ZPL (cf. Fig.~\ref{Fig5}(b)). The Debye–Waller factor increases inversely proportional to the dopant atom size, which reduces distortions of the lattice~\cite{orphal2025coherent, thiering2018ab, bradac2019quantum}. 

All group-IV vacancy centers possess a similar energy-level structure for their electronic ground and excited states~\cite{harris2024coherence, orphal2025coherent}. Both states form spin-doublet and orbital-doublet manifolds that are governed by the same types of interactions, though with different coupling strengths~\cite{orphal2025coherent}. For these vacancy centers, the intrinsic Hamiltonian is dominated by spin-orbit coupling and the Jahn-Teller effect \cite{hepp2014electronic, neu2013low, muller2014optical}. These interactions lift the fourfold degeneracy of both the ground and excited states, splitting them into a lower branch (LB) and an upper branch (UB). In addition, external magnetic fields give rise to Zeeman interactions, while lattice strain couples to the orbital degrees of freedom in both electronic states. Fig.~\ref{Fig4}(b) depicts the electronic energy level structure in detail~\cite{hepp2014electronic, thiering2018ab}.

The electron spin-$\frac{1}{2}$ states are denoted by $|\uparrow\rangle$ and $|\downarrow\rangle$, corresponding to the spin projections $|m_S=+1/2\rangle$ and $|m_S=-1/2\rangle$, respectively. The orbital states are described in the $|e_{+}\rangle$ and $|e_{-}\rangle$ basis, for which the spin-orbit interaction Hamiltonian is diagonal~\cite{beukers2025quantum, thiering2018ab, hepp2014electronic}. Spin-orbit coupling is a relativistic interaction that couples the electron spin to its orbital angular momentum and becomes stronger for the heavier group-IV elements, leading to characteristic splittings $\Delta_{\mathrm{GS}}$ and $\Delta_{\mathrm{ES}}$ in the ground and excited states, respectively.

The Jahn-Teller effect originates from a spontaneous symmetry breaking of the electronic wavefunction and affects the Hamiltonian in two ways~\cite{beukers2025quantum, thiering2018ab, hepp2014electronic}. First, it partially quenches the orbital angular momentum, effectively reducing the spin-orbit coupling strength $\lambda$. Second, it induces a linear displacement of the nuclei, which enters the Hamiltonian in the same manner as static lattice strain~\cite{beukers2025quantum}.

Strain in the diamond lattice couples directly to the electronic orbital states and can be decomposed into two symmetry components with distinct effects. Longitudinal strain produces a uniform energy shift of all levels within either the ground or the excited state; since this shift differs between the two manifolds, it modifies the optical transition frequency. Transverse strain, in contrast, mixes the orbital states and splits the lower and upper branches by introducing an off-diagonal term in the spin-orbit Hamiltonian~\cite{stramma2024tin, beukers2025quantum}.

Combining all contributions—including spin-orbit coupling \cite{beukers2025quantum, stramma2024tin, meesala2018strain}, transverse strain \cite{stramma2024tin, beukers2025quantum}, the electronic-spin Zeeman effect \cite{stramma2024tin, beukers2025quantum}, and orbital angular momentum \cite{stramma2024tin, karapatzakis2024microwave, beukers2025quantum}—yields the total Hamiltonian of the group-IV vacancy center. The energy separation between the lower and upper branches plays a central role in determining the coherence properties of the group-IV vacancy centers and is given by~\cite{stramma2024tin, beukers2025quantum}

\begin{equation}
\Delta_{\mathrm{GS(ES)}} = \sqrt{\lambda^2 + 4\epsilon_{GS(ES)}^2},
\end{equation}

The qubit can be defined by considering the two levels of the lower branch of the ground state (cf. Fig.~\ref{Fig4}(b)), which exhibits the longest coherence time. The coherence of both the spin qubit and the optical transition in group-IV vacancy centers is strongly temperature dependent, as decoherence is dominated by phonon interactions in the diamond lattice~\cite{stramma2024tin, beukers2025quantum, Jahnke2015phonons, Torun2023optical}. The leading mechanism is a direct single-phonon process, in which a phonon excites the electron from the lower branch (LB) to the upper branch (UB) at a rate $\Gamma_{+}$, followed by relaxation back to the LB through phonon emission at a rate $\Gamma_{-}$:

\begin{align}
\Gamma_{+} &= 2\pi \chi \rho \Delta_{\mathrm{GS}}^{3} n(\Delta_{\mathrm{GS}},T), \\
\Gamma_{-} &= 2\pi \chi \rho \Delta_{\mathrm{GS}}^{3} [n(\Delta_{\mathrm{GS}},T)+1],
\end{align}

where $\chi$ is the electron-phonon coupling strength~\cite{beukers2025quantum}, $\rho$ is the phonon density of states~\cite{beukers2025quantum}, and

\begin{equation}
n(\Delta_{\mathrm{GS}},T) = \frac{1}{e^{\Delta_{\mathrm{GS}}/k_{\mathrm{B}}T}-1}
\end{equation}

is the Bose-Einstein occupation factor. In the low-temperature regime ($k_{\mathrm{B}}T \ll \Delta_{\mathrm{GS}}$)~\cite{stramma2024tin, beukers2025quantum}, the excitation rate satisfies $\Gamma_{+} \ll \Gamma_{-}$~\cite{stramma2024tin, beukers2025quantum}, such that the orbital relaxation time

\begin{equation}
T_{1,\mathrm{orbit}} = \frac{1}{\gamma_{+}}
\end{equation}

sets the fundamental limit for all phonon-induced decoherence processes. Each phonon scattering event fully randomizes the spin phase, leading to a spin dephasing time $T_{2,\mathrm{spin}}^{*} = T_{1,\mathrm{orbit}}$~\cite{beukers2025quantum}. Spin relaxation occurs more slowly, since not every orbital transition induces a spin flip. This is quantified by the spin cyclicity $\eta_{\mathrm{spin}}$, giving~\cite{stramma2024tin, beukers2025quantum}

\begin{equation}
T_{1,\mathrm{spin}} = \eta_{\mathrm{spin}}\, T_{1,\mathrm{orbit}},
\end{equation}

with $\eta_{\mathrm{spin}} \approx 1200$ for aligned magnetic fields~\cite{stramma2024tin, beukers2025quantum}. Figure~\ref{Fig4}(c) shows the simulated and experimentally measured orbital coherence time with respect to the ground state splitting and temperature for negatively-charged SiV~\cite{rogers2014all, pingault2017coherent, sohn2018controlling}, GeV~\cite{siyushev2017optical}, and SnV~\cite{trusheim2020transform, guo2023microwave, rosenthal2023microwave} color defects in diamond. For spin-photon interface applications, the coherence time of the optical transition must exceed the excited-state lifetime. At low temperatures, optical dephasing is dominated by the ground state due to its smaller splitting~\cite{beukers2025quantum}, resulting in

\begin{equation}
T_{2,\mathrm{optical}}^{*} = T_{1,\mathrm{orbit}}.
\end{equation}

Phonon-induced decoherence can be mitigated by increasing the energy splitting $\Delta_{\mathrm{GS}}$~\cite{beukers2025quantum}. Heavier group-IV elements naturally provide larger splittings, and an applied strain can further enhance $\Delta_{\mathrm{GS}}$, enabling higher operating temperatures~\cite{beukers2025quantum, Klotz2025ultra}. An alternative approach is to reduce the phonon density of states by confining defects in nanodiamonds, although this strategy is practically viable only for silicon-vacancy centers due to phonon wavelength constraints~\cite{beukers2025quantum, Klotz2022prolonged}. Phononic crystals where the phononic transition frequencies overlap with the bandgap, such that transitions are inhibited, have also been shown for reducing thermal decoherence \cite{Kuruma2025controlling}.

The coherence of electronic and nuclear spin qubits is summarized in Table~\ref{Table-1.1}. Owing to their long coherence times, NV centers and other negatively charged group-IV defect centers in diamond are promising candidates for use as quantum nodes, where the electronic spin can function as a communication qubit and the nuclear spin as a memory qubit (nuclear spin coherence time is longer than the electronic spin coherence time).

\section{Recent progress in diamond color center platforms for metropolitan-scale quantum networking}

Although diamond color centers have emerged as highly promising qubit systems for quantum technologies \cite{chatterjee2021semiconductor, zhang2020material, atature2018material}, the scalability of diamond-based quantum devices remains limited by diamond’s incompatibility with mature micro- and nanofabrication processes~\cite{schroder2016quantum, katsumi2025recent, wan2020large}. To address this challenge, significant efforts have focused on the hybrid integration of diamond quantum systems with complementary material platforms that offer advanced photonic and electronic functionality~\cite{schroder2016quantum, katsumi2025recent, wan2020large, chen2024scalable, codreanu2025diamond}.

Silicon nitride has become a widely adopted photonic platform in the visible wavelength range, benefiting from commercial-scale fabrication across multiple foundries~\cite{chen2024scalable, pelucchi2022potential}. Its broad transparency window, ultra-low propagation loss~\cite{poon2018integrated}, and mature processing technologies make it particularly well-suited for passive photonic components such as waveguides~\cite{chen2024scalable, senichev2022silicon}, resonators~\cite{jun2023ultrafast, luo2023recent}, and filters~\cite{bryan2023biosensing}. Nevertheless, the lack of strong optical nonlinearities in silicon nitride limits its applicability for active photonic elements.

In parallel, emerging platforms such as aluminum nitride~\cite{wan2020large}, aluminum gallium nitride \cite{Gundogdu2025algan, Gundogdu2026hetero}, and lithium niobate~\cite{pan2023perspective, assumpcao2024thin, shang2023inverse, hu2025integrated} have attracted considerable interest due to their combination of low optical loss with substantial piezoelectric and nonlinear optical coefficients. These properties enable a wide range of active functionalities, including electro-optic modulation, fast optical switching, and frequency conversion~\cite{hu2025integrated, assumpcao2024thin}. Hybrid integration of diamond nanophotonic devices with such photonic integrated circuits—using approaches such as transfer printing or pick-and-place assembly—offers a viable route toward scalable, modular quantum architectures~\cite{wan2020large, nickerson2014freely}.

Hybrid quantum–photonic platforms enable the co-integration of quantum emitters~\cite{kim2020hybrid}, optical control fields, photon routing, spectral filtering, and detection within a single chip or compact package~\cite{hausmann2012integrated, chakravarthi2023hybrid, majumder2022engineering, abulnaga2025design, butcher2020high, schrinner2020integration}. This level of integration is a key requirement for the realization of large-scale~\cite{knaut2024entanglement}, multi-qubit quantum networks~\cite{wan2020large, chung2025interqnet}. While many of the required components are already well established in classical integrated photonics, their direct translation to quantum systems remains challenging, primarily due to optical loss constraints. In particular, fully on-chip photon routing and control typically require cascaded high-speed switches and filters, which can introduce prohibitive losses for single-photon applications ~\cite{esmann2024solid}.

Efficient interfacing between diamond quantum devices and photonic integrated circuits is another critical challenge~\cite{katsumi2025recent, wan2020large, chen2024scalable, kim2020hybrid}. Experimental demonstrations using evanescent-field coupling in pick-and-place architectures have reported diamond-to-PIC coupling efficiencies of approximately $40\%$ with aluminum nitride~\cite{pasini2024nanophotonics, wan2018efficient, wan2020large} and up to $90\%$ with lithium niobate~\cite{pasini2024nanophotonics, riedel2023efficient, desiatov2019ultra}, highlighting both the promising and material dependence of hybrid approaches. Additional challenges arise in coupling visible-wavelength photons from chip-based platforms to free-space optics or optical fibers, where fabrication techniques and commercial solutions remain significantly less mature than in the telecom wavelength regime.

Figure~\ref{Fig1} provides a schematic overview of a large-scale quantum network and illustrates the concept of a quantum node based on a hybrid-integrated device~\cite{pasini2024nanophotonics}. In such a node, a diamond nanophotonic structure hosting a color center with an associated nuclear-spin register is interfaced with integrated photonic and electronic circuitry for coherent control of both the spin and photonic qubits. Recent progress in scalable hybrid nanophotonic architectures, strategies for increasing processor size, and quantum frequency conversion techniques has enabled demonstrations of metropolitan-scale quantum networking. These developments represent important milestones toward the realization of distributed quantum information processing based on diamond defect centers~\cite{nickerson2014freely, ruf2021quantum}.

\subsection{Required building blocks for metropolitan-scale quantum networks}
\subsubsection{Communication and memory qubits}

\begin{figure*}[]
\centering
\includegraphics[width=\textwidth]{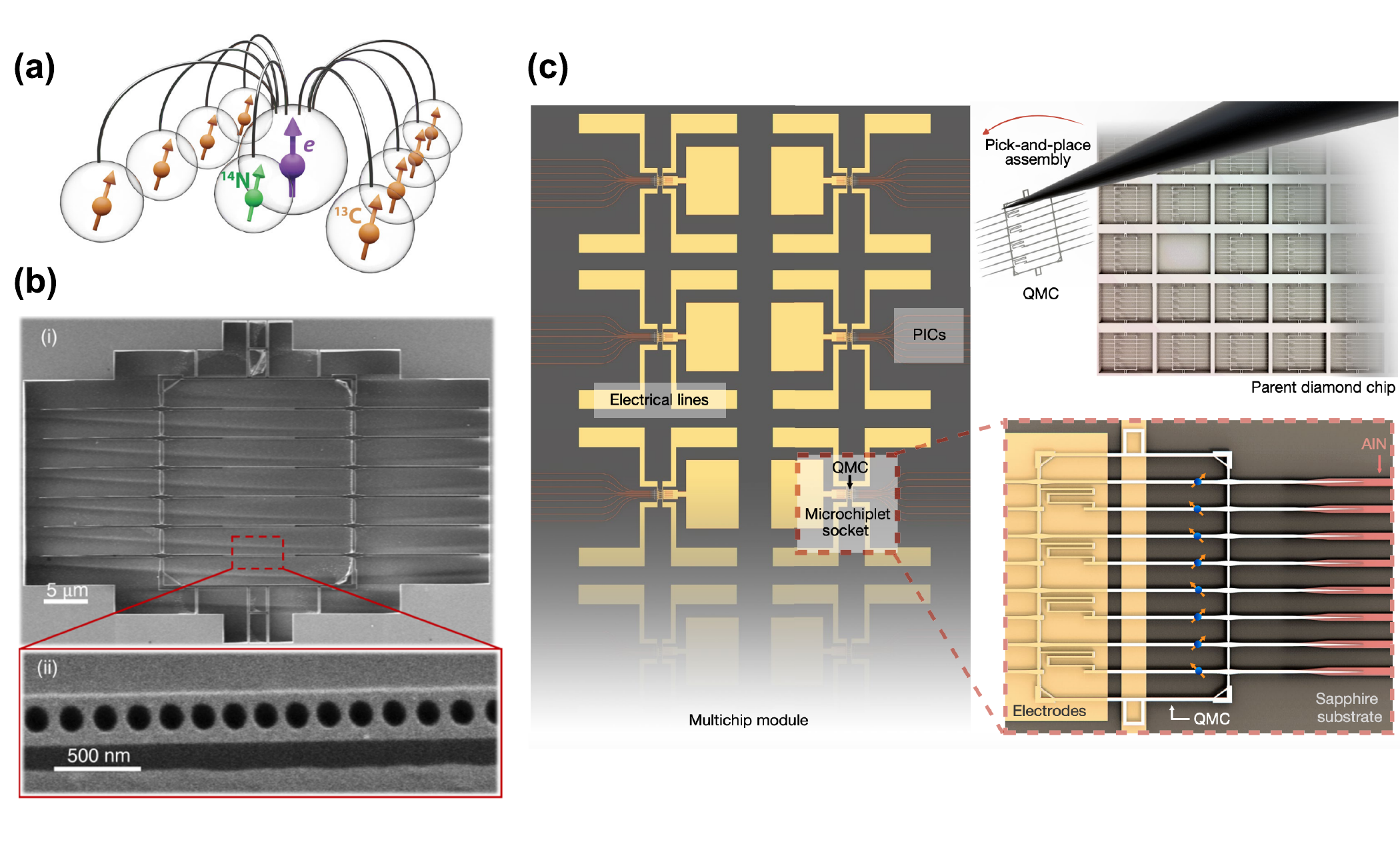}
\caption{(a) Illustration of the ten-qubit register. The electron spin of a single NV center in diamond serves as the central qubit and is coupled, via hyperfine interactions, to the intrinsic $^{14}$N nuclear spin and eight surrounding $^{13}$C nuclear spins. This figure is adapted from Ref.~\cite{bradley2019ten}. (b) Scanning electron microscope (SEM) images of (i) six fabricated diamond photonic crystal (PhC) cavities integrated with silicon nitride (SiN) waveguides within a quantum microchiplet (QMC), and (ii) a magnified view of the central region of an individual cavity. This figure is adapted from Ref.~\cite{chen2024scalable}. (c) The figure illustrates a scalable approach for integrating diamond color centers with nanophotonic components. In this hybrid platform, subcomponents are fabricated and optimized separately prior to final assembly, maximizing yield, device size, and overall performance. Pre-screened QMCs are subsequently transferred from the parent diamond chip into a host platform using a pick-and-place technique, enabling efficient photonic interfacing as well as electrical access for spectral tuning and spin-qubit control. This figure is adapted from Ref.~\cite{wan2020large}.}
\label{Fig6}
\end{figure*}

Extending diamond defect center-based two-node quantum networks to multi-node architectures requires additional qubit resources capable of storing quantum information while new entanglement links are established. Such quantum memories~\cite{bersin2024telecom, lei2023quantum} are essential for implementing entanglement distillation~\cite{kalb2017entanglement}, purification~\cite{bennett1996purification, deutsch1996quantum}, and quantum repeater protocols~\cite{briegel1998quantum, borregaard2019quantum}, which collectively enhance the size (cf. Fig.~\ref{Fig6}(a)), speed, and communication range of quantum networks. The diamond-based color-defect center (i.e., NV center) is particularly well suited for these tasks, owing to its excellent coherence and control properties (cf. Table~\ref{Table-1.1})\cite{orphal2025coherent}. At cryogenic temperatures ($\sim$4\,K), the electronic and nuclear spin coherence times reach $T_2^{\mathrm{(e)}} \approx 1.5$\,s and $T_2^{\mathrm{(n)}} \approx 1$\,min \cite{pezzagna2021quantum}, respectively. In addition, quantum gate operations are fast, with single-qubit electron-spin gate times below 10\,ns~\cite{fuchs2009gigahertz} and two-qubit gate times of approximately 700\,ns~\cite{rong2015experimental}. High-fidelity operation has also been demonstrated, with single-qubit gate fidelities approaching 99.995$\%$~\cite{rong2015experimental}, two-qubit electron–electron spin gate fidelities exceeding 97$\%$~\cite{yamamoto2013strongly}, and electron–nuclear spin gate fidelities of about 99.2$\%$~\cite{rong2015experimental}. Furthermore, long-distance entanglement fidelities of approximately 92$\%$ have been achieved over 1\,km distance~\cite{hensen2015loophole}.

A key requirement for an effective quantum memory is robustness against the entanglement-generation protocol, which typically must be repeated many times before success is achieved. In this context, carbon ($^{13}$C) nuclear spins in diamond constitute particularly attractive quantum memories compatible with diamond defect center–based quantum networks~\cite{bradley2019ten}. Their weak hyperfine coupling to the color-defect center electronic spin enables quantum states to be stored with minimal loss during repeated attempts at spin–photon entanglement generation~\cite{knaut2024entanglement}. Theoretical studies indicate that, under realistic experimental coupling parameters~\cite{beukers2024remote}, quantum information encoded in such nuclear spins can be preserved over hundreds of entanglement-generation cycles~\cite{blok2015towards}. These results underscore the strong potential of weakly coupled nuclear spins in diamond as robust quantum memories~\cite{bradley2019ten, knaut2024entanglement} and highlight their critical role in the realization of scalable quantum networks based on solid-state defect centers. For other color centers, the electronic spin qubit and nearby $^{13}$C nuclear spin qubits can similarly be employed as communication and memory qubits, respectively, using appropriate combinations of MW and RF pulse sequences~\cite{bradley2019ten, knaut2024entanglement, orphal2025coherent, siyushev2017optical, karapatzakis2024microwave, rosenthal2023microwave}.

\subsubsection{Scalable hybrid integrated quantum device}

A central challenge in quantum networking is the coherent transfer of quantum states between different physical modalities, most notably between flying qubits and stationary quantum memories~\cite{yang2016high, iuliano2024qubit, tsurumoto2019quantum}. One promising approach to this problem relies on spin–photon interfaces that combine solid-state spin qubits—such as color centers in diamond—with nanophotonic structures~\cite{chen2024scalable, wan2020large, Schroder2017Scalable}. While high-fidelity spin–photon interactions have been demonstrated in isolated devices~\cite{mouradian2017rectangular, ding2024high, li2015coherent, pregnolato2024fabrication, joe2024high, parker2024diamond}, scaling these interfaces to the large numbers required for practical quantum repeaters remains an outstanding challenge~\cite{almutlaq2026foundry, kim2020hybrid, li2024heterogeneous}.

Recent progress has addressed this challenge through the use of SnV centers coupled to integrated diamond cavities assembled into quantum microchiplets (QMCs)~\cite{chen2021polarization}. Figure~\ref{Fig6}(b) illustrates a hybrid architecture in which a diamond QMC is integrated with a silicon nitride (SiN) photonic integrated circuit (PIC), enabling optical and electrical addressing of individual color centers. The oxide-clad photonic chip features openings that allow low-loss SiN waveguides to evanescently couple to individual diamond QMC channels. Each channel contains a diamond nanophotonic cavity coupled to an input–output waveguide mode~\cite{mouradian2017rectangular}. In this platform, Purcell enhancement was demonstrated for four cavity-coupled SnV systems within a six-channel QMC, achieving an average Purcell factor of $\simeq 7$. Furthermore, a separate cavity QMC was heterogeneously integrated into a SiN PIC, enabling on-chip spectroscopy via waveguide coupling.

Figure~\ref{Fig6}(c) presents a complementary approach based on high-yield heterogeneous integration of diamond QMCs—waveguide arrays hosting highly coherent color centers—onto photonic integrated circuits. Using this technique, a defect-free \(128\)-channel array of GeV and SiV centers was integrated onto an aluminum nitride (AlN) PIC~\cite{wan2020large}. Photoluminescence spectroscopy revealed long-term stability and narrow average optical linewidths of 54\,MHz (146\,MHz) for GeV (SiV) emitters, approaching their respective lifetime-limited linewidths of 32\,MHz (93\,MHz). Inhomogeneous broadening of individual optical transitions was mitigated through in situ tuning over a range of 50\,GHz without linewidth degradation. The ability to assemble large numbers of nearly indistinguishable and tunable artificial atoms within phase-stable photonic integrated circuits represents a critical advancement toward multiplexed quantum repeaters and scalable, general-purpose quantum processors.

\subsubsection{Quantum frequency conversion}

In long-distance implementations, optical fiber loss (cf. Fig.~\ref{Fig2}(d)) becomes the dominant factor limiting the achievable entanglement generation rate~\cite{ruf2021quantum, wolfowicz2021quantum, atature2018material}. Many leading solid-state platforms for quantum processor nodes emit photons in the visible spectral range (cf. Table~\ref{Table-1.1}). However, fiber attenuation at these wavelengths severely constrains operation to distances of only a few kilometers (cf. Fig.~\ref{Fig2}(d)).

By contrast, optical fiber loss exhibits a global minimum near 1550\,nm wavelength, corresponding to the telecommunications C-band (cf. Fig.~\ref{Fig2}(d)). Accessing this low-loss transmission window requires converting the emission wavelength of photons generated at quantum processing nodes into the telecom band. Quantum frequency conversion (QFC) provides a means to achieve this while preserving the quantum correlations or entanglement between the photons and the underlying qubit states~\cite{kumar1990quantum, huang1992observation}. As such, state-preserving frequency conversion of single photons represents a key element of the quantum interface for interconnecting remote quantum information processing nodes, enabling long-range, low-loss transmission through optical fiber~\cite{knaut2024entanglement, stolk2024metropolitan, bersin2024telecom}.

\begin{figure*}[]
\centering
\includegraphics[width=\textwidth]{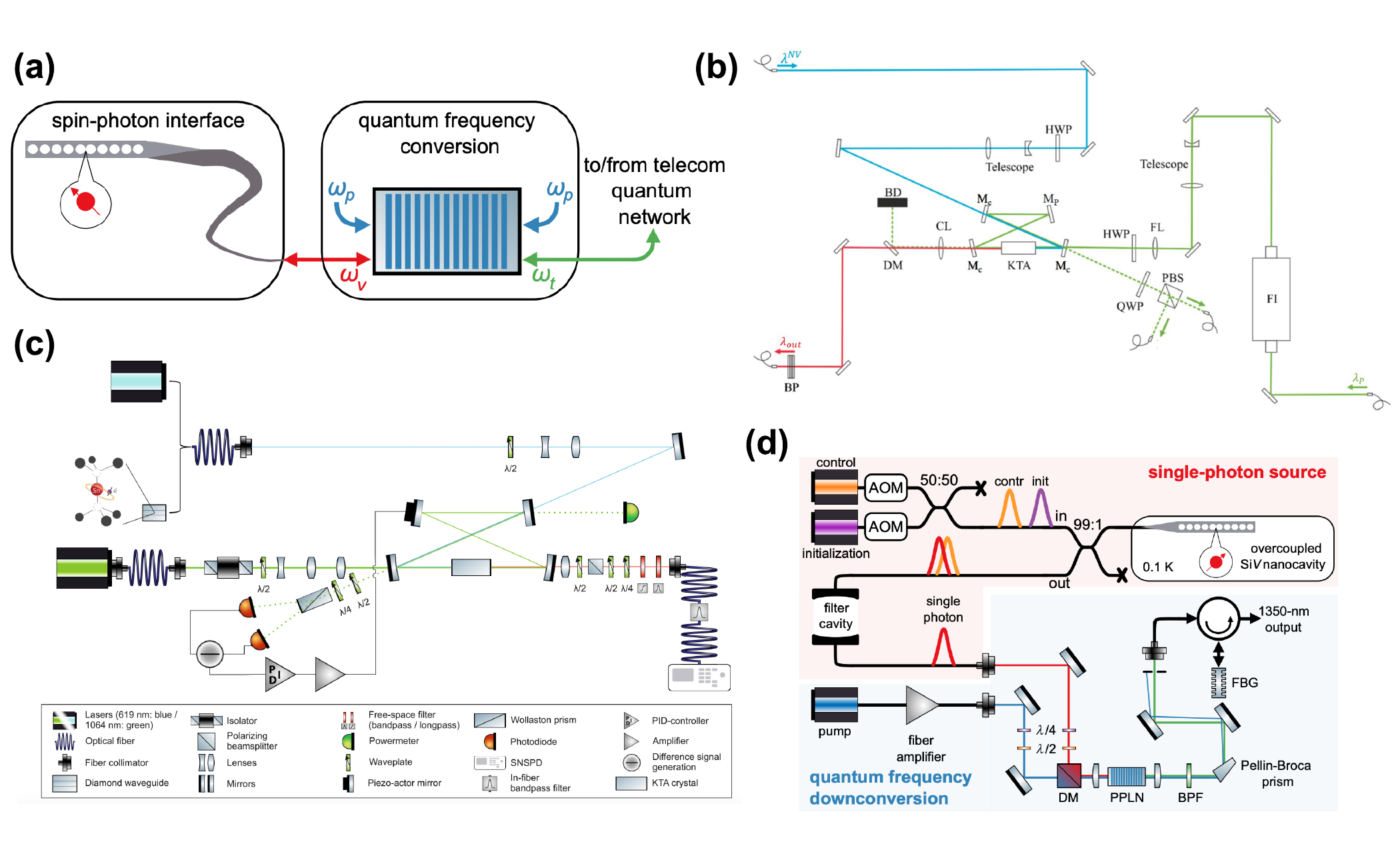}
\caption{(a) Spin–photon quantum network node based on a single defect center in diamond (e.g., SiV) coupled to a nanophotonic cavity. Visible emission at frequency $\omega_{\mathrm{v}}$ is converted to a telecom frequency $\omega_{\mathrm{t}}$ via quantum frequency conversion (QFC) using a pump at $\omega_{\mathrm{p}}$. This figure is adapted from Ref.~\cite{bersin2024telecom}. (b) Cavity-enhanced QFC using a bulk KTA nonlinear crystal inside an actively stabilized optical cavity, enabling low-noise conversion of photons emitted by NV centers. This figure is adapted from Ref.~\cite{geus2024low}. (c) Free-space QFC of single photons generated by a waveguide-integrated SnV center, with filtered telecom output coupled into fiber for detection. This figure is adapted from Ref.~\cite{brevoord2025quantum}. (d) Waveguide-based QFC of single photons emitted by an SiV center using a periodically poled lithium niobate (PPLN) waveguide, preserving single-photon purity and indistinguishability. This figure is adapted from Ref.~\cite{bersin2024telecom}.}
\label{Fig7}
\end{figure*}

Recent experimental advances have demonstrated highly efficient and ultra-low-noise QFC platforms capable of converting visible photons to telecom wavelengths without degrading their quantum properties~\cite{bersin2024telecom, geus2024low, iuliano2024qubit, brevoord2025quantum}. One prominent approach employs an intermediate-frequency pump laser resonantly enhanced within an actively stabilized optical cavity containing a bulk monocrystalline nonlinear crystal. Using photons emitted by NV center qubits as a testbed, this system achieved an external conversion efficiency of up to $43\%$, accompanied by an exceptionally low noise photon rate of approximately $2\,\mathrm{s}^{-1} \mathrm{pm}^{-1}$ ($17\,\mathrm{s}^{-1}\mathrm{GHz}^{-1}$)~\cite{geus2024low,stolk2024metropolitan}. This corresponds to a reduction in noise by nearly two orders of magnitude compared with earlier implementations based on periodically poled waveguide crystals. The tunability of the converted output wavelength further enables the generation of indistinguishable telecom photons from spectrally distinct network nodes, highlighting the suitability of this approach for scalable fiber-based quantum networks~\cite{geus2024low, raymer2012manipulating, stolk2024metropolitan}. A schematic of this QFC implementation is shown in Fig.~\ref{Fig7}(b).

A related challenge arises for diamond SnV centers, whose native emission wavelength of 619\,nm is incompatible with metropolitan-scale fiber networks. This limitation has been addressed through the demonstration of efficient and low-noise QFC from 619\,nm to the telecom S-band at 1480\,nm. In this implementation~\cite{brevoord2025quantum}, visible photons are combined with 1064\,nm pump light inside an actively stabilized cavity incorporating a bulk monocrystalline potassium titanyl arsenate (KTA) crystal. Internal and external conversion efficiencies of $(48 \pm 3)\%$ and $(28 \pm 2)\%$~\cite{brevoord2025quantum}, respectively, were achieved, along with a spectrally flat noise photon rate of $(2.2 \pm 0.9)$ counts s$^{-1}$pm$^{-1}$ over a 40\,GHz bandwidth. Notably, the conversion efficiency remained above $80\%$ of its peak value over a tuning range of approximately 70\,GHz~\cite{brevoord2025quantum}. Successful conversion was confirmed using photons generated by a waveguide-integrated SnV center, with the converted telecom photons preserving the characteristic excited-state lifetime of the emitter (cf. Fig.~\ref{Fig7}(c)).

Beyond SnV-based systems, bidirectional and low-noise QFC has also been demonstrated for SiV centers in diamond. In this case~\cite{bersin2024telecom}, single photons emitted in the visible band were converted to the telecom O-band while maintaining their nonclassical character, as evidenced by a second-order autocorrelation $g^{(2)}(0) < 0.1$~\cite{bersin2024telecom}, as well as high photon indistinguishability with a visibility of $V = (89 \pm 8)\%$~\cite{bersin2024telecom}. Collectively, these results establish quantum frequency conversion as a mature and scalable interface technology that enables solid-state quantum memories based on diamond color centers to directly interconnect with existing telecom-band infrastructure. Such advances represent a critical step toward metropolitan-scale fiber-based quantum networks and, ultimately, a global quantum internet \cite{chung2025interqnet, bersin2024development}. A schematic illustration of this QFC approach is shown in Fig.~\ref{Fig7}(d).

Beyond these QFC demonstrations, recent progress in second-harmonic generation (SHG) in diamond nanocavities~\cite{flaagan2025optical} has revealed that the effective second-order susceptibility, $\chi^{(2)}$, strongly depends on the electronic configuration of defects such as NV-centers. In particular, photoionization from the negatively charged to the neutral state modifies $\chi^{(2)}$, leading to quenching of SHG under green illumination. Controlled green excitation therefore enables optical switching of the device nonlinearity. This charge-state control of the defect centers opens new opportunities for second-order nonlinear processes in diamond, including frequency conversion, optical modulation, and all-optical switching, thereby expanding the capabilities of diamond photonics.

\subsection{Experimental demonstrations of large-scale quantum networks with diamond defect centers}

Recent advances have addressed this challenge through the demonstration of a two-node quantum network based on multi-qubit registers formed by silicon-vacancy (SiV) centers embedded in nanophotonic diamond cavities and interfaced with a telecommunications fiber network. In this approach \cite{knaut2024entanglement}, remote entanglement is mediated by cavity-enhanced spin–photon interactions between the electronic spin qubits of the SiV centers and optical photons.

High-fidelity entanglement between distant nodes is achieved using serial, heralded spin–photon entangling gate operations implemented with time-bin photonic qubits. Long-lived nuclear spin qubits associated with the $^{29}$Si isotope function as quantum memories, enabling entanglement storage over second-long timescales while also providing integrated error detection during the entanglement generation process. By incorporating efficient, bidirectional quantum frequency conversion of photonic communication qubits to telecommunications wavelengths near 1350~nm \cite{bersin2024telecom}, this platform demonstrates entanglement of two nuclear spin memories over both 40~km spools of low-loss optical fiber and a 35~km deployed fiber loop in an urban metropolitan environment. Together, these results mark a significant step toward practical quantum repeaters and scalable quantum networking architectures.

Experimentally, the electronic spin coherence times of the two network nodes are measured to be 125~$\mu$s and 134~$\mu$s, respectively, while the $^{29}$Si nuclear spin exhibits a storage lifetime exceeding 2~s. The reported fidelities for electron–electron and nuclear–nuclear spin entanglement are approximately 0.86 and 0.77, respectively. Nanophotonic cavities play a crucial role in enhancing the spin–photon interaction, thereby improving the overall efficiency of the entanglement protocols. The measured single-photon cooperativity is 12.4 for node~A and 1.5 for node~B. Analysis of the electron-spin–dependent cavity reflection spectra enables direct extraction of the spin–photon cooperativity, providing a quantitative measure of the emitter–cavity coupling strength.

\begin{figure*}[]
\centering
\includegraphics[width=\textwidth]{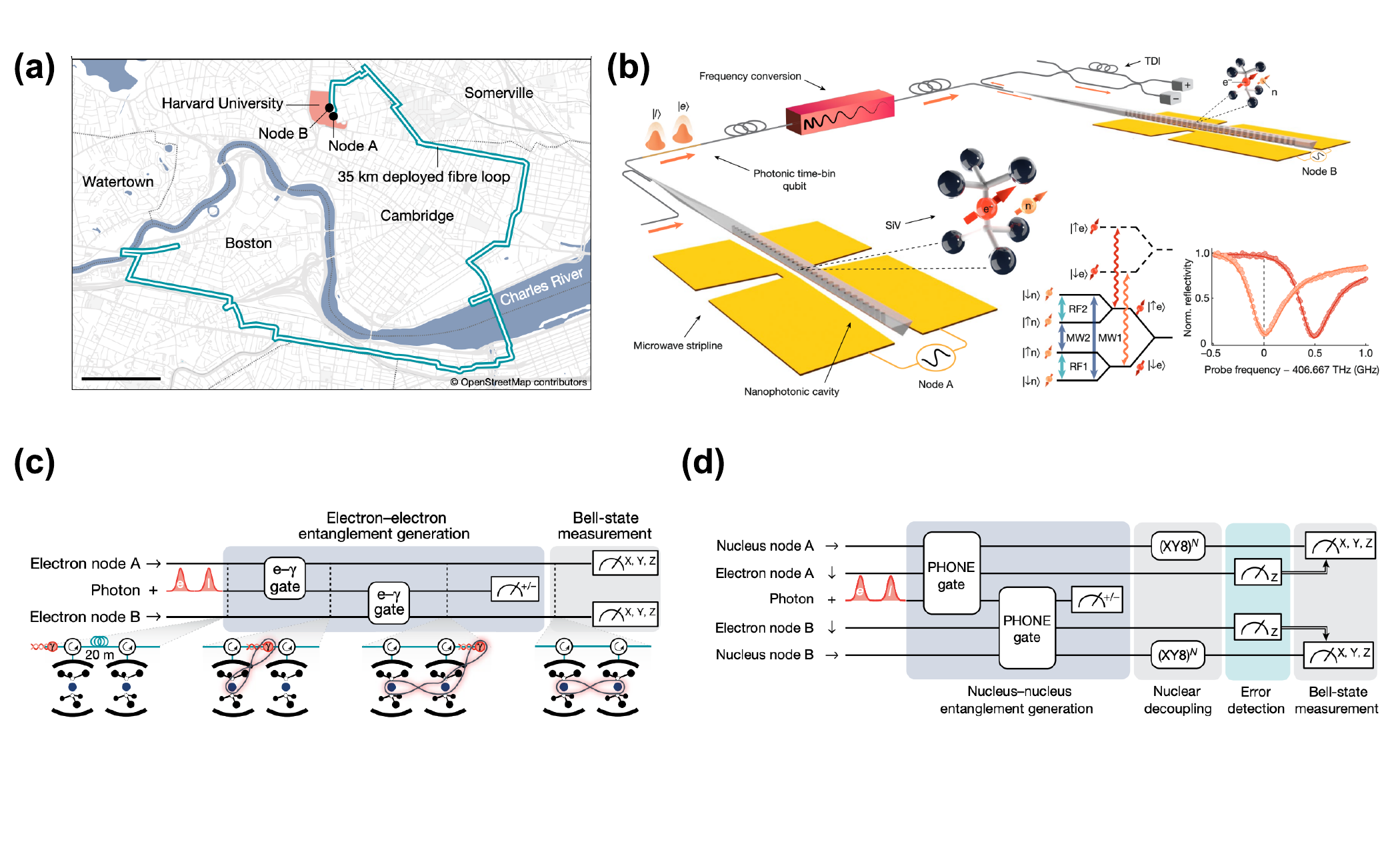}
\caption{(a) Deployed telecom fiber link connecting nodes A and B, spanning 35\,km across the greater Boston metropolitan area.
(b) Experimental setup with individual $^{29}$SiV centers embedded in nanophotonic cavities and operated in separate cryostats below 200\,mK. The nodes are connected either by a short fiber link with active frequency tuning or by a long telecom fiber link using quantum frequency conversion (QFC). Microwave and radio-frequency control provides access to the electronic and nuclear spin transitions of the $^{29}$SiV center. (c) A remote entanglement protocol in which time-bin photonic qubits sequentially interact with the electron spins at nodes A and B, generating heralded electronic Bell states via photon measurement. (d) Nuclear-nuclear entanglement generated by sequentially entangling a time-bin photonic qubit with the $^{29}$Si nuclear spins at nodes A and B using two PHONE gates. This figures are adapted from Ref-\cite{knaut2024entanglement}.}
\label{Fig8}
\end{figure*}

Figure~\ref{Fig8} illustrates the implementation of a metropolitan-scale quantum network deployed in the Boston area~\cite{bersin2024development}.

\begin{figure*}[]
\centering
\includegraphics[width=\textwidth]{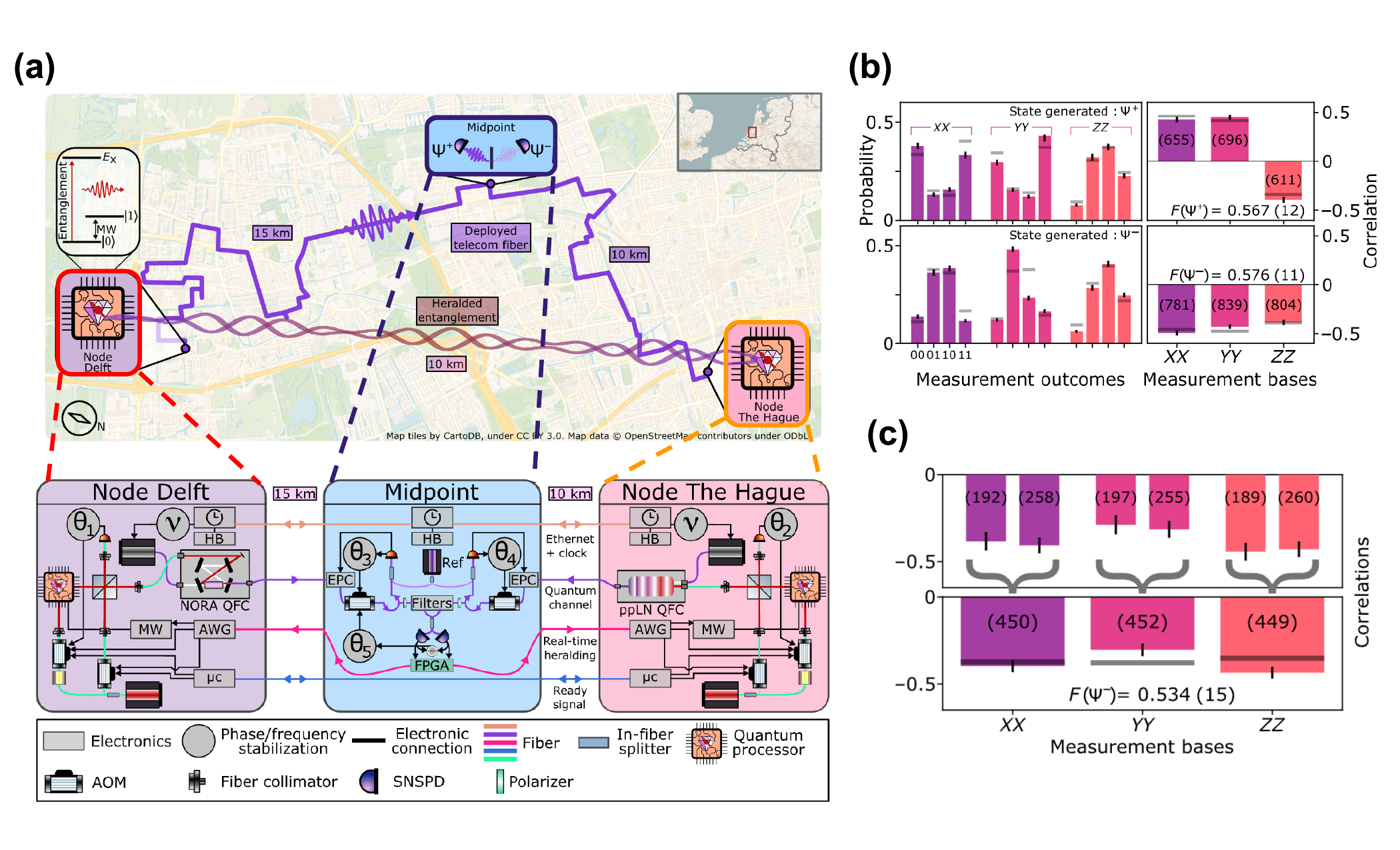}
\caption{(a) Schematic of a metropolitan-scale quantum link connecting two independently operated NV-center-based quantum network nodes in Delft and The Hague via deployed optical fiber. Qubits are encoded in the electronic ground-state spin and manipulated with microwave pulses, while spin-selective optical transitions at 637 nm enable entanglement generation and readout. (b) Representative qubit-qubit correlation measurements for different heralding detector outcomes, together with the corresponding extracted state fidelities and theoretical predictions. (c) Correlation measurements under full heralding conditions, where both detector signatures project the remote nodes onto the same Bell state, demonstrating heralded remote entanglement at metropolitan scale. This figures are adapted from Ref-\cite{stolk2024metropolitan}.}
\label{Fig9}
\end{figure*}

Another significant advance was reported by Ronald Hanson’s group at TU Delft (Ref.~\cite{stolk2024metropolitan}), where heralded entanglement was demonstrated between two independently operated quantum network nodes separated by 10\,km (cf. Fig.~\ref{Fig9}). The two nodes, hosting diamond spin qubits based on the ground-state electronic spin of NV centers, were connected to a central midpoint station via 25\,km of deployed optical fiber. To mitigate photon loss in the fiber, the native qubit photons were converted to the telecom L-band using quantum frequency conversion~\cite{stolk2022telecom, geus2024low}, and the link was embedded in an extensible, phase-stabilized architecture that enables a loss-resilient single-click entanglement protocol~\cite{bose1999proposal, cabrillo1999creation}. By exploiting the full heralding capabilities of the network together with real-time feedback on the long-lived qubits, the experiment demonstrated the preparation of a predefined entangled state at the nodes, independent of the specific heralding detection outcome. This work represents the first demonstration of heralded qubit–qubit entanglement at the metropolitan scale, achieving post-selected and fully heralded entanglement fidelities of 0.576 and 0.534~\cite{stolk2024metropolitan}, respectively, using both cavity-enhanced (NORA, based on a KTA crystal) and periodically poled lithium niobate waveguide QFC devices~\cite{stolk2022telecom, geus2024low}.

\section{Challenges and prospective solutions for diamond defect-based qubits}

This section reviews the key challenges associated with diamond-based defect centers in the context of quantum networking. It first addresses limitations arising from the optical and spin properties of diamond defect centers from a materials perspective, followed by challenges encountered in metropolitan-scale quantum network demonstrations \cite{atature2018material, ruf2021quantum}. Central issues include entanglement fidelity, the probability of successful entanglement generation, and the indistinguishability of photons emitted by distinct defect centers \cite{stolk2024metropolitan, bersin2024telecom, iuliano2024qubit, knaut2024entanglement}. The scalability of quantum nodes based on diamond defect centers, incorporating optical and electronic on-chip control, is then discussed. Finally, the section surveys potential mitigation strategies, including the use of high-cooperativity optical cavities to enhance entanglement rates, recent advances in diamond nanophotonics, methods to improve photon indistinguishability, and the integration of additional nuclear spin qubits to enable scalable quantum node architectures.

\subsection{Challenges related to the diamond color center based qubit systems and proposed solutions \label{section:challenges}}

While the NV center has served as a central platform for quantum network demonstrations in diamond, several intrinsic limitations currently constrain its performance. In particular, only about $3\%$ of the NV center’s optical emission occurs in the ZPL \cite{atature2018material, ruf2021quantum}, and photon collection efficiencies are limited due to the high-refractive index of the diamond \cite{hepp2014electronic}. As a result, spin-photon entanglement generation rates are typically restricted to below \(100\) Hz \cite{ruf2021quantum, pompili2021realization, humphreys2018deterministic}. In addition, the NV center’s permanent electric dipole moment makes it highly sensitive to charge fluctuations in its local environment, leading to pronounced spectral diffusion on short timescales, especially for centers located within a few micrometers of a surface \cite{ruf2019optically, atature2018material, ruf2021quantum}.

\begin{figure*}[]
\centering
\includegraphics[width=\textwidth]{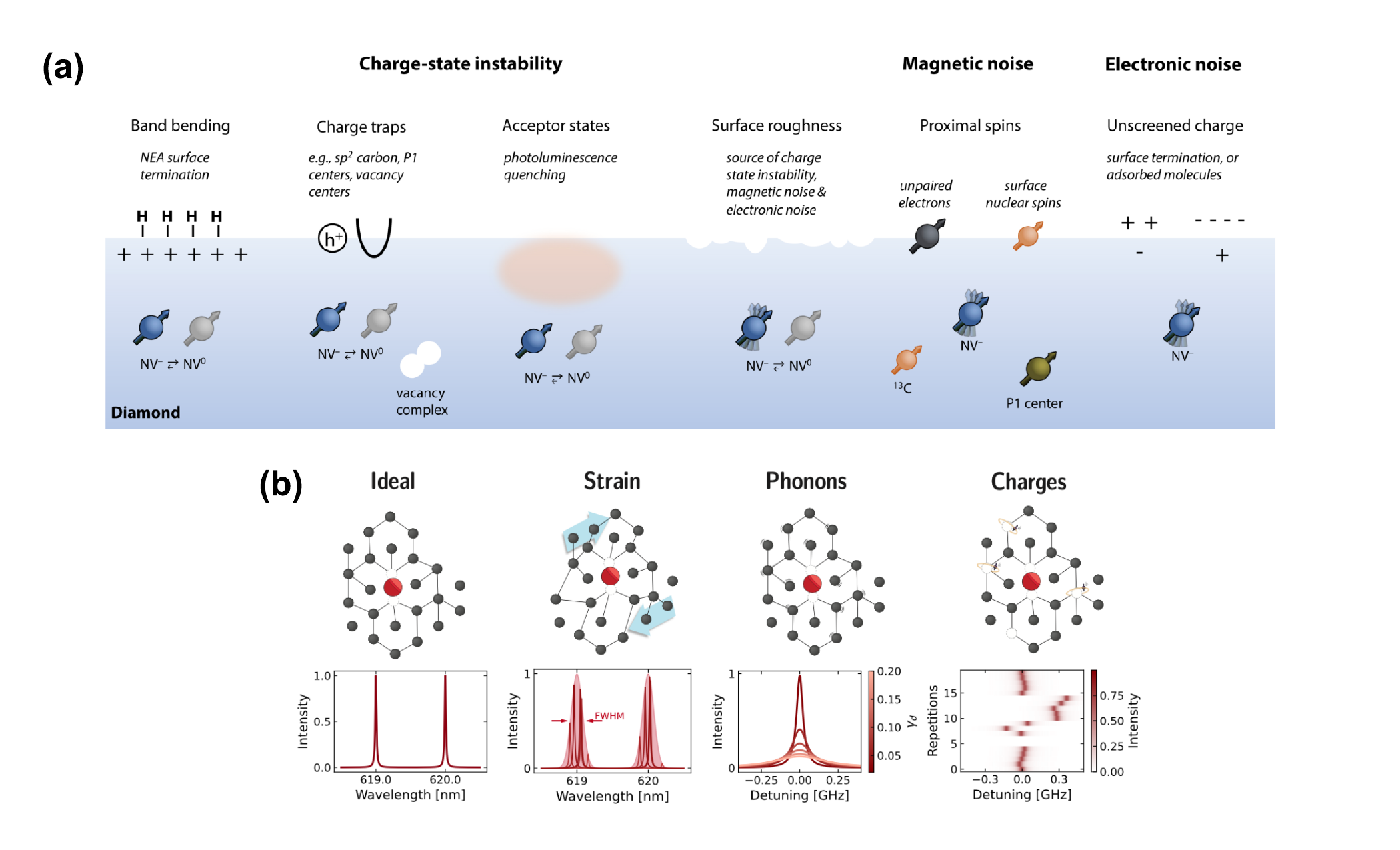}
\caption{(a) Overview of charge-state instability and decoherence mechanisms for near-surface NV centers in diamond. The figure summarizes the dominant surface- and near-surface-related processes that affect the optical stability and spin coherence of NV centers. Charge-state instability arises from surface-induced band bending, charge traps (e.g., sp$^2$ carbon, P1 centers, and vacancy-related defects), and acceptor states that lead to photoluminescence quenching and NV to NV$^{0}$ conversion and vice versa. Additional contributions stem from surface roughness, which can introduce charge-state fluctuations as well as magnetic and electronic noise. Magnetic decoherence is primarily caused by proximal paramagnetic species, including unpaired surface electrons, surface nuclear spins, and nearby bulk spins such as $^{13}$C and P1 centers. Electronic noise further originates from unscreened surface charges associated with surface termination or adsorbed molecules. Together, these effects represent key limitations for near-surface NV centers in nanophotonic devices and quantum networking applications. This figure is adapted from Ref-\cite{janitz2022diamond}. (b) Group-IV vacancy (like SnV) center in the diamond crystal and the dominant sources of optical instability. In an ideal, strain-free crystal environment, the group-IV vacancy (like SnV) center exhibits well-defined optical transition wavelengths corresponding to the C and D lines. Spatial variations in the local strain of the diamond host shift the transition frequencies from emitter to emitter, giving rise to an inhomogeneous distribution of optical resonances. Phonon-induced pure dephasing broadens the intrinsic optical linewidth beyond the lifetime limit. Fluctuations in the local charge environment manifest as spectral diffusion and abrupt spectral jumps during repeated excitation, and can also lead to fluorescence intermittency due to charge-state switching. This figure is adapted from Ref-\cite{pasini2024nanophotonics}.}
\label{Fig10}
\end{figure*}

These effects have so far prevented access to the regime of coherent cooperativity \(C_{\mathrm{coh}} > 1\) \cite{ruf2021cavity, ruf2021quantum}, despite reports of Purcell factors as high as \(70\) in small mode-volume photonic crystal cavities \cite{faraon2012coupling}. The microscopic origins of surface-induced noise remain an active area of investigation. Recent experiments probing the interaction between near-surface NV center spins and their local electrostatic environment have identified unwanted signals arising from so-called dark electron spins at the diamond surface (cf. Fig.~\ref{Fig10}(a)) \cite{janitz2022diamond, zvi2025engineering}. A surface-modification approach \cite{zvi2025engineering, yu2026engineering} based on passivating the diamond with a thin titanium oxide (TiO\(_2\)) layer has been shown to substantially reduce the density of these dark spins. The observed reduction—from typical values of approximately \(2{,}000~\mu\mathrm{m}^{-2}\) to below the detection limit of NV centers (\(\sim 200~\mu\mathrm{m}^{-2}\))-is consistent with the electronic structure of the diamond-TiO\(_2\) interface and results in a twofold increase in the Hahn-echo coherence time of near-surface NV centers. Combined with proposals for active optical stabilization to suppress spectral diffusion, such advances may help revive the prospects of NV centers in nanophotonic architectures \cite{zvi2025engineering}.

Another important aspect is the controlled creation of shallow, optically stable NV centers. Techniques such as electron irradiation and laser writing can produce optically coherent NV centers \cite{chen2019laser}; however, they rely on native nitrogen impurities to form NV defects through vacancy recombination, limiting precise spatial control over defect placement. Deterministic positioning is particularly critical for maximizing Purcell enhancement \cite{fehler2019efficient}, as NV centers must be located at antinodes of the cavity field. Although ion implantation can provide such spatial precision, NV centers created via this method have been shown to exhibit increased optical linewidths compared with NV centers formed during growth, even after extensive high-temperature annealing to repair lattice damage. Consequently, achieving coherent cooperativity and thereby approaching the near-deterministic spin-photon interface regime remains a significant challenge for NV center-based nanophotonic systems \cite{ruf2021quantum, schroder2016quantum}.

The inherent first-order insensitivity to electric fields exhibited by group-IV color centers in diamond has enabled the observation of optical transitions that approach the lifetime limit, even within extensively nanofabricated photonic structures \cite{orphal2025coherent, thiering2018ab, rugar2020narrow, sipahigil2014indistinguishable, evans2016narrow, muller2014optical}. To date, low-temperature coupling of group-IV color centers to photonic crystal cavities has been realized for both SiV and SnV centers, with the SiV system achieving coherent cooperativities exceeding 100. This strong light-matter interaction permits spin-state-dependent photon reflection, allowing high-fidelity single-shot spin readout with fidelities approaching 99.9$\%$ \cite{ruf2021quantum}.

\begin{figure*}[]
\centering
\includegraphics[width=\textwidth]{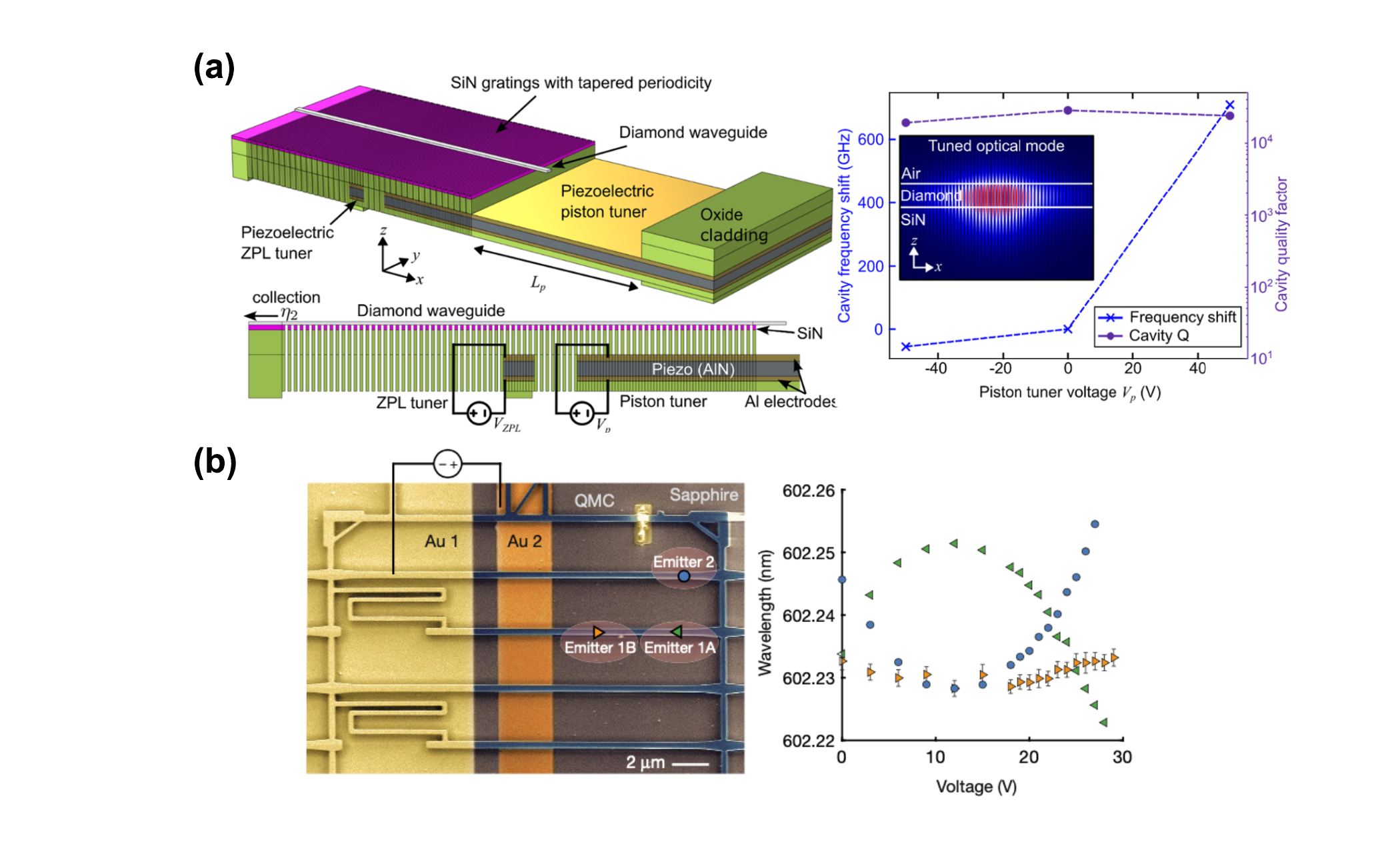}
\caption{Integrated strain tuning of hybrid diamond nanophotonic cavities. (a) Schematic of a hybrid cavity strain-tuning platform incorporating two independent piezoelectric actuators: a zero-phonon-line (ZPL) tuner and a piston tuner, controlled by voltages \(V_{\mathrm{ZPL}}\) and \(V_p\), respectively. The piston tuner enables cavity-mode tuning over a range of up to 760\,GHz (from \(-50\) to \(50\)\,V) while preserving the cavity quality factor. Inset: simulated cavity electric-field magnitude in the XZ plane at \(V_p = 0\). This figure is adapted from Ref-\cite{greenspon2025designs}. (b) SEM image of the quantum micro-chiplet. Strain-dependent optical tuning of emitters 1A, 1B, and 2 is shown, enabling spectral overlap within a single waveguide (emitters 1A and 1B at 24.5\,V) and between different waveguides (emitters 2 and 1A/1B at 2\,V and 12\,V, respectively). This figure is adapted from Ref-\cite{wan2020large}.}
\label{Fig11}
\end{figure*}

A critical requirement for remote entanglement generation is the ability to tune multiple group-IV color centers, located on separate chips, into mutual resonance. Emission frequency tuning has been demonstrated using strain (cf. Fig.~\ref{Fig11}) \cite{wan2020large, machielse2019quantum, Sohn2018}, electric fields \cite{de2021investigation, aghaeimeibodi2021electrical}, and Raman-based approaches \cite{sipahigil2016integrated, pingault2014all}. However, these techniques have so far been limited to tuning individual emitters or multiple emitters within the same structure. Both strain- and electric-field-based tuning modify the electronic orbitals of the color center, breaking inversion symmetry and leading to increased optical linewidths and enhanced spectral diffusion \cite{aghaeimeibodi2021electrical, de2021investigation}, likely due to greater sensitivity to environmental charge noise. Experimental evidence suggests that strain tuning offers a larger tuning range for comparable levels of line broadening when compared to electric-field tuning. Raman-based schemes, while avoiding some of these issues, are restricted to emission-based entanglement protocols.

Although near-lifetime-limited linewidths have been observed for group-IV color centers in nanophotonic devices, experiments frequently report residual spectral shifts and charge instabilities, resulting in linewidth broadening by several times the lifetime limit as well as significant local strain fields. Dynamic strain tuning can mitigate slow spectral diffusion occurring on timescales of seconds, but it remains challenging to reduce homogeneous linewidth broadening on sub-microsecond timescales, which directly impacts the achievable coherent cooperativity.

Group-IV color centers are typically formed through high-energy ion implantation followed by annealing to generate vacancy complexes and partially repair implantation-induced lattice damage. However, such treatments do not fully restore the pristine diamond lattice. Promising alternative approaches include low-energy shallow implantation combined with diamond overgrowth, as well as strategies that pair low-density ion implantation with electron irradiation to minimize lattice damage. When combined with deliberate control of the diamond Fermi level, these techniques offer potential pathways toward improving the optical stability and coherence of group-IV color centers.

The optical performance of group-IV vacancy centers is strongly influenced by the quality of the surrounding diamond crystal. Several dominant noise mechanisms arise from the crystal environment and impact the optical transitions. Figure-\ref{Fig10}(b) shows the effect of strain, phononic interaction, and presence of charge on the spectral properties of the group-IV color centers. Local strain variations, originating from crystallographic defects, dislocations, implantation damage, or nanofabrication-induced stress, lead to spatially varying transition frequencies across different emitters, resulting in inhomogeneous broadening. Additionally, phonon-induced pure dephasing modifies the spectral content of emitted photons, broadening the optical transition beyond the lifetime limit. Finally, despite their reduced sensitivity to electric fields, group-IV vacancy centers remain susceptible to fluctuations in the local charge environment, which manifest as shot-to-shot variations in transition frequency, commonly referred to as spectral diffusion.

An additional challenge associated with group-IV color centers is the direct microwave driving of spin transitions \cite{rosenthal2023microwave, karapatzakis2024microwave, Pieplow2024efficient}. Group-IV centers exhibit an inherent resistance to spin mixing due to strong spin-orbit coupling.
As the size of the implanted atom increases, the spin-orbit interaction gets stronger, leading to an even weaker coupling to the magnetic dipoles of the group-IV centers and microwave fields.
One approach to enhance spin mixing is to pre-select emitters under intrinsic strain or by applying external strain. However, this typically leads to a degradation of the optical quality of the emitted photons.

\subsection{Challenges related to diamond qubit-based metropolitan-scale quantum links}

Low photon collection efficiencies are a critical challenge for scaling up quantum networks. 
These problems are amplified at the metropolitan scale, as photons must be converted to the telecom band, and state-of-the-art conversion efficiencies range from 10\% \cite{bersin2024telecom} to 50\% \cite{geus2024low}. Strong pump power is required to enhance the frequency conversion efficiency; however, it also increases the noise level due to Raman scattering and weakly phase-matched parametric down-conversion processes that arise in the nonlinear medium under intense pump power \cite{chung2025interqnet}. A recent advancement has been reported in the visible-to-telecom frequency conversion of weak coherent pulses at the single-photon level \cite{wengerowsky2025quantum}.
The additional loss channel has an inhibiting effect on the achievable entanglement rates. 
As more and more network nodes are included in the non-deterministic entanglement attempts, the entanglement rates will drop further into impractical values.

Similarly, reported entanglement fidelities (88\% for SiV \cite{knaut2024entanglement}, 53\% for NV \cite{stolk2024metropolitan}) over metropolitan networks are significantly lower than unity.
Since infidelities accumulate across each quantum node, scaling to additional nodes appears mathematically unfeasible without improving on the state-of-the-art values.

One should also note that setting up a node, even with cutting-edge equipment, is a tremendous experimental challenge that only a few groups worldwide can achieve.
Therefore, scaling up also requires more easily accessible and operable equipment, such as cryogenics with optical access and electrical infrastructure, and logic-compatible experimental electronics that are synchronized over all the quantum nodes.
In addition, implementing the high-rate single-click protocols requires phase- and polarization-stabilized fiber links between the quantum nodes.

Furthermore, fabricating samples that host emitters meeting the required conditions is not a widespread capability. 
An ideal sample for scaling up would involve emitters with long electron spin coherence times, nuclear spin coupling, both controllable by microwave delivery systems, emission maintaining lifetime-limited linewidths, and strong coupling to nanocavities. 

Finally, it is still an open question, even if all the quantum nodes can successfully generate photons on the telecom band that successfully interfere, how all these photons could be routed around the world.
Although the existing telecom fiber network infrastructure offers an opportunity, it is still unclear how a quantum internet can be integrated without hindering classical communication.


\subsection{Recent developments related to the efficiency of quantum nodes and proposed solutions}

Microwave control of emitters is typically implemented by golden striplines within integrated applications.
But, this comes with the trade-off that the delivered powers heat the sample and cause additional decoherence on the controlled emitter.
Niobium deposition on the diamond sample was successfully used to fabricate superconducting striplines \cite{karapatzakis2024microwave}, thereby resulting in significantly lower heating.

Since group-IV centers are not naturally occurring in diamond, they must be created by high-energy atom implantation. 
This results in the generation of many lattice defects that, through ionization, trap charges and contribute to the electric-field noise around the emitters.
Although conventional annealing steps can help to produce emitters with lifetime-limited emission, it still requires a preselection from hundreds to have negligible spectral diffusion.
It turns out, however, increasing the annealing temperature to 2100 $^{\circ}$C under high pressure ($>$ 6 GPa) to prevent graphitization, can deterministically create optically stable narrow linewidth emitters with a mean linewidth of 34 MHz and a central frequency distribution full-width-half-maximum of 3.9 GHz \cite{narita2023multiple}.

Suspended structures like nanocavities and waveguides can be fabricated using quasi-isotropic etching \cite{Khanaliloo2015single} or Faraday cages \cite{Latawiec2016faraday}.
The resulting structures, however, suffer from high sidewall roughness. 
As an alternative method, it was shown that starting with diamond membranes with the thickness of the intended structure and then applying isotropic etching results in cavities that perform better, with a record-high quality factor of $1.8 \times 10^5$ \cite{ding2024high}.
The so-called 'smart-cut' method involves implanting helium atoms and annealing to generate a graphitized layer at a target depth, which can be removed with acids. 
The remaining diamond layer can then be transferred to another substrate using PDMS stamps \cite{Guo2021tunable}.

One additional challenge in increasing the nanofabrication output is preparing the masks for the structures. 
The conventional method for patterning diamond masks with electron beam lithography is a lengthy process.
Recently, 750x750 µm silicon masks were developed independently in foundries and transferred onto diamonds.
Using these masks, 120 QMCs composed of 15 photonic structures were fabricated.
This development removes one of the more difficult steps in patterning and enables a faster production of different nanostructures \cite{almutlaq2026foundry}.

Another development in hetero-integration of diamond structures is the integration of QMCs into CMOS chips.
On this chip, spin preparation and spectroscopic control of different emitters have been demonstrated.
This constitutes an important step in the development of quantum chips with photonic and electronic integration \cite{Li2024heterogenous}.

\subsection{Future directions in scaling and applications}

The largest network demonstrated so far with diamond-based color centers is a three-node network  \cite{pompili2021realization}. 
There, a central node was used to entangle two outer nodes in the chain.
The next natural step in implementing an extended network would be to connect a central node to more than two nodes, enabling entanglement swapping and quantum teleportation in multiple directions.


It is worth noting here that a similar extension in scaling has been proposed in the context of heterogeneous networks \cite{chung2025interqnet}. 
There are multiple proposed applications for quantum networks that utilize the inherent communication security based on the principles of quantum mechanics \cite{Bozzio2025quantum}.
One such example is quantum dice rolling, which involves a complete random ordering of users across different nodes \cite{Aharon2010quantum} and can be used in protocols that require randomized decision-making.
Another example is enabling a voting process without depending on a central election authority \cite{Centrone2022quantum}.
An additional demonstration could be the use of a multi-qubit photonic state as a quantum token to authenticate the identities of different nodes \cite{Strocka2025secure}.

One of the important prospects of quantum networks is the ability to distribute computations for enhanced quantum information processing. 
Selected proof-of-principle applications on small networks, especially those with a clear quantum advantage, would constitute significant demonstrations. 
Such an application was exhibited on a two-node network by implementing a blind computing application, in which the server solved a problem for the client without enough information to reconstruct the calculation later \cite{Wei2025universal}. 
Another application of an advanced quantum computation prospect is the preparation of cluster states for measurement-based computation.

Entanglement can increase sensitivity in many measurements, such as atomic clocks, magnetometers, gravimeters, gyroscopes \cite{Huang2024entanglement}.
A recent example uses a remote entangled SiV pair to detect a phase difference in an interferometric measurement \cite{Stas2026entanglement}.
Demonstrations of other applications in which networks deliver measurable improvements in metrological performance would be highly valuable.

How to combine the advantages of different quantum platforms to enhance the performance of quantum nodes is an active area of research. 
One such proposal is to combine the bright, optically desirable properties of quantum dots for entangled photon pair generation with group-IV color centers as the memory qubit to develop hybrid quantum repeaters \cite{Strocka2025hybrid}.

Developing quantum devices compatible with large-scale manufacturing and exhibiting consistent, dependable performance is a general objective for all aspirations of quantum technologies.
Widely and commercially available photonic chips with electronic integration, hosting pre-characterized optically and spin-coherent emitters, would significantly accelerate scientific and technological development.

\subsection{Perspective}
\paragraph{Architectures for diamond-based quantum networks.}

As highlighted in the introduction, diamond-based quantum memories can, in principle, be used in all major repeater architectures, including detection-in-midpoint, sender-receiver, and source-in-midpoint schemes. Although comparing these approaches is challenging, Refs.~\cite{omlor2025pulseReflection, beukers2024remote} provide a unified framework for evaluating protocols across a broad range of cavity and interface parameters, including present-day and projected cooperativities. They also discuss several photonic encodings, including time-bin, polarization, dual-rail, and frequency encoding. For color centers in diamond, time-bin encoding appears especially natural, although frequency and polarization encoding are also possible, depending on the interface and protocol design \cite{beukers2024remote,omlor2025pulseReflection}.
Encoding strategies such as time-bin encoding, which are compatible with reflection-based phase gates, are among the most promising approaches. Reflection-based phase gates can, in principle, approach unit efficiency for entanglement distribution \cite{Strocka2025hybrid}, making them attractive for hybrid schemes that combine bright quantum-dot photon sources with diamond color-center memories \cite{Strocka2025hybrid}.

Comparing diamond-based quantum networks with other platforms is intrinsically difficult because quantum memories can operate at very different wavelengths and temperatures and can also have very different spatial requirements, which are important for scalability. A useful overview of fiber-based platforms is given in Ref.~\cite{loock2020modules}, while Ref.~\cite{farre2026qpufReview} summarizes representative state-of-the-art memory figures of merit across wavelength ranges and operating temperatures. In this broader landscape, diamond color centers do not offer the longest storage times, but they compare favorably in interface efficiency and compatibility with nanophotonic integration \cite{bhaskar2020experimental,bopp2024sawfish}. This distinguishes them from many atomic and ionic platforms, where excellent coherence is often accompanied by more demanding optical interfacing and reduced compatibility with scalable on-chip nanophotonics.

A common figure of merit for comparing repeater architectures is the secret-key rate, since it captures both the final state quality and the usable communication rate. However, secret-key rate alone is often insufficient for a fair comparison, because the relevant resource cost can differ substantially between protocols. Depending on the architecture, one may need to account for the number of photons sent per successful link, the number of communication and memory qubits, ancillary qubits, multiplexing overhead, gate and measurement fidelities, photon source brightness, collection and detection efficiencies, classical communication latency, repeater spacing, total node count, and the relative hardware cost of different node types \cite{borregaard2020oneWayRepeater, wo2023resource, Strocka2025hybrid}. Accordingly, several recent works introduce secret-key-rate benchmarks with explicit cost functions or resource-normalized performance measures \cite{wo2023resource, borregaard2020oneWayRepeater}.

Hybrid proposals also combine ensemble memories, such as rubidium-based or atomic frequency comb (AFC)-type systems, with single-spin photon transducers to unite long-lived storage with efficient spin-photon gates and strong photonic nonlinearities \cite{Gu2025HybridQuantumRepeaters}. In this context, diamond color centers are particularly attractive because their nanosecond optical lifetimes enable fast optical cycling and thus potentially higher interface rates. At the same time, as discussed in Sec.~\ref{section:challenges}, diamond still faces important challenges, most notably spectral diffusion and material-dependent inhomogeneity, so its present-day optical stability generally remains below that of the cleanest atomic and ionic systems \cite{orphalKobin2023nvNanostructures}.

\section{Conclusion}

Diamond defect centers, with their long spin coherence times and favorable optical properties, are well-suited as candidates for quantum network nodes. Progress from laboratory-scale experiments to metropolitan-scale demonstrations has established their potential for scalable quantum technologies. Nonetheless, significant challenges remain, particularly in photon indistinguishability, efficient light–matter interactions, entanglement success rates, and large-scale integration. Advances in nanophotonics and cavity quantum electrodynamics offer promising pathways to overcome these limitations, and continued development in these directions is expected to enable practical large-scale quantum networks.

\section*{Acknowledgment}

KS acknowledges funding from the Department of Science and Technology (DST), Indian National Quantum Mission (NQM). PK acknowledges funding from the U.S. National Science Foundation, Award No. 2231388, through the Expanding Capacity in Quantum Information Science and Engineering (ExpandQISE) program. TS acknowledges funding from the European Research Council (ERC), project HyperGraph, 101171255, and the German Federal Ministry of Research, Technology and Space (BMFTR), project Quantenrepeater.Net (QR.N), 16KIS2185. AM acknowledges fellowship from I-Hub Chanakya Doctoral Fellowship. TS and GP acknowledge funding from the German Federal Ministry of Research, Technology, and Space (BMFTR), project QPIS.2, KIS6GCQ020.


\section*{Author contribution}

All authors contributed to the preparation of the manuscript.

\section*{Competing interests}

The authors declare no competing interests.



%

\end{document}